\documentclass[aps,prx,reprint,amsmath,amssymb,twocolumn, superscriptaddress, floatfix]{revtex4}

\usepackage{graphicx}
\usepackage{dcolumn}
\usepackage{bm}
\usepackage{braket}
\usepackage{hyperref}
\usepackage{xcolor}
\usepackage{upgreek}
\begin{document}

\preprint{APS/123-QED}

\title{High-fidelity dispersive spin sensing in a tuneable unit cell \\ of silicon MOS quantum dots}

\author{Constance Lainé}
\email[]{constance.laine.19@ucl.ac.uk}
\affiliation{Quantum Motion, 9 Sterling Way, London N7 9HJ, United Kingdom}
\affiliation{London Centre for Nanotechnology, UCL, 17-19 Gordon St,~London~WC1H~0AH,~United~Kingdom}

\author{Giovanni A. Oakes}%
\affiliation{Quantum Motion, 9 Sterling Way, London N7 9HJ, United Kingdom}
\author{Virginia Ciriano-Tejel}
\email[]{virginia@quantummotion.tech}
\affiliation{Quantum Motion, 9 Sterling Way, London N7 9HJ, United Kingdom}

\author{Jacob~F.~Chittock-Wood}
\affiliation{Quantum Motion, 9 Sterling Way, London N7 9HJ, United Kingdom}
\affiliation{London Centre for Nanotechnology, UCL, 17-19 Gordon St,~London~WC1H~0AH,~United~Kingdom}

\author{Lorenzo~Peri}
\affiliation{Quantum Motion, 9 Sterling Way, London N7 9HJ, United Kingdom}
\affiliation{Cavendish Laboratory, University of Cambridge, J. J. Thomson Avenue, Cambridge CB3 0HE, United Kingdom}

\author{Michael A. Fogarty}
\affiliation{Quantum Motion, 9 Sterling Way, London N7 9HJ, United Kingdom}
\author{Sofia~M.~Patomäki}
\affiliation{Quantum Motion, 9 Sterling Way, London N7 9HJ, United Kingdom}
\author{Stefan Kubicek}
\affiliation{IMEC, Kapeldreef 75, 3001 Leuven, Belgium}
\author{David F. Wise}
\affiliation{Quantum Motion, 9 Sterling Way, London N7 9HJ, United Kingdom}
\author{Ross C. C. Leon}
\affiliation{Quantum Motion, 9 Sterling Way, London N7 9HJ, United Kingdom}
\author{M. Fernando Gonzalez-Zalba}
\affiliation{Quantum Motion, 9 Sterling Way, London N7 9HJ, United Kingdom}
\author{John J. L. Morton}
\affiliation{Quantum Motion, 9 Sterling Way, London N7 9HJ, United Kingdom}
\affiliation{London Centre for Nanotechnology, UCL, 17-19 Gordon St,~London~WC1H~0AH,~United~Kingdom}

\date{\today}
\begin{abstract}
Metal-oxide-semiconductor (MOS) technology is a promising platform for developing quantum computers based on spin qubits. Scaling this approach will benefit from compact and sensitive sensors that minimize constraints on qubit connectivity while being industrially manufacturable. Here, we demonstrate a compact dispersive spin-qubit sensor, a single-electron box (SEB), within a bilinear unit cell of planar MOS quantum dots (QDs) fabricated using an industrial grade 300 mm wafer process. By independent gate control of the SEB and double-quantum-dot tunnel rates, we optimize the sensor to achieve a readout fidelity of 99.92\% in 340$~\upmu$s (99\% in 20$~\upmu$s), fidelity values on a par with the best obtained with less compact sensors. Furthermore, we develop a Hidden Markov Model of the two-electron spin dynamics that enables a more accurate calculation of the measurement outcome and hence readout fidelity. Our results show how high-fidelity sensors can be introduced within silicon spin-qubit architectures while maintaining sufficient qubit connectivity as well as providing faster readout and more efficient initialisation schemes.
\end{abstract}

\maketitle

An essential component within a scalable quantum processor unit (QPU) is a qubit measurement device that combines high readout fidelity with a minimised physical footprint~\cite{Laucht_2021}. 
For semiconductor spin qubits, readout is usually performed by mapping the spin to a charge state detected using a nearby charge sensor such as the single-electron-transistor (SET)~\cite{fulton1987observation, Angus2007, Kranz2020},
which has set the standard for fast high-fidelity readout (see Supplementary Note~\ref{app:fid_benchmark}). 
Recently, there have been demonstrations of a more compact type of charge sensor, the radio-frequency (RF) single-electron box (SEB), which consists of a single charge reservoir coupled to a quantum dot (QD), whose impedance is measured via RF reflectometry~\cite{Gonzalez2015, House2016, Urdampilleta2019, Ansaloni2020, Ciriano-Tejel2021, oakes2023fast, Niegemann2022, Hogg2023Single-ShotSensor}. In addition to being more compact, SEBs have been predicted to offer sensitivities approaching those of SETs, assuming suitable resonator and device optimisation~\cite{Oakes2023}. 
%

One key advantage of semiconductor spin qubits is their potential compatibility with high-yield fabrication processes of the semiconductor industry~\cite{Gonzalez2021a,DeMichielis2023}, particularly for architectures based on existing transistor technology, such as planar silicon metal-oxide-semiconductor (MOS)~\cite{veldhorst2017silicon, Li2018}. In this technology, spin readout has typically been done using SETs due to their large resistance swings, resulting in high sensitivity, despite their larger footprint. 
%
Here, we study whether SEB charge sensors in MOS spin qubit devices can offer readout speeds and fidelities comparable to SETs, while offering the potential for more scalable architectures. We use a QPU unit cell comprising one row of QDs hosting spin qubits, with a parallel row that includes an SEB. Our approach is to leverage the electrical tunability afforded by multiple overlapping gates, available within the planar MOS fabrication process. By adjusting the potential of barrier gates between adjacent QDs we optimise the signal-to-noise ratio (SNR) and readout fidelity.
%
Next, we introduce a Hidden Markov Model (HMM) that considers the full readout dynamics, accounting for three possible initial states of the double quantum dot spin system. 
This approach offers two improvements over methods which simplify the system by considering only two states (typically one triplet and the singlet): i) more accurately estimating the true readout fidelity and ii) better classifying single shot traces than traditional threshold methods. Together, these device-level and analytical advances allow us to optimise spin relaxation rates for maximum fidelity in parity readout --- which we find exceeds 99.9\%. 
Finally, The HMM also enables us to distinguish three of the four two-spin states in a single shot, offering the potential for more efficient initialisation schemes~\cite{Philips2022}.

\section{\label{sec:level2} QPU unit cell with a scalable sensor}

\begin{figure}
    \centering
    \includegraphics[width=\linewidth]{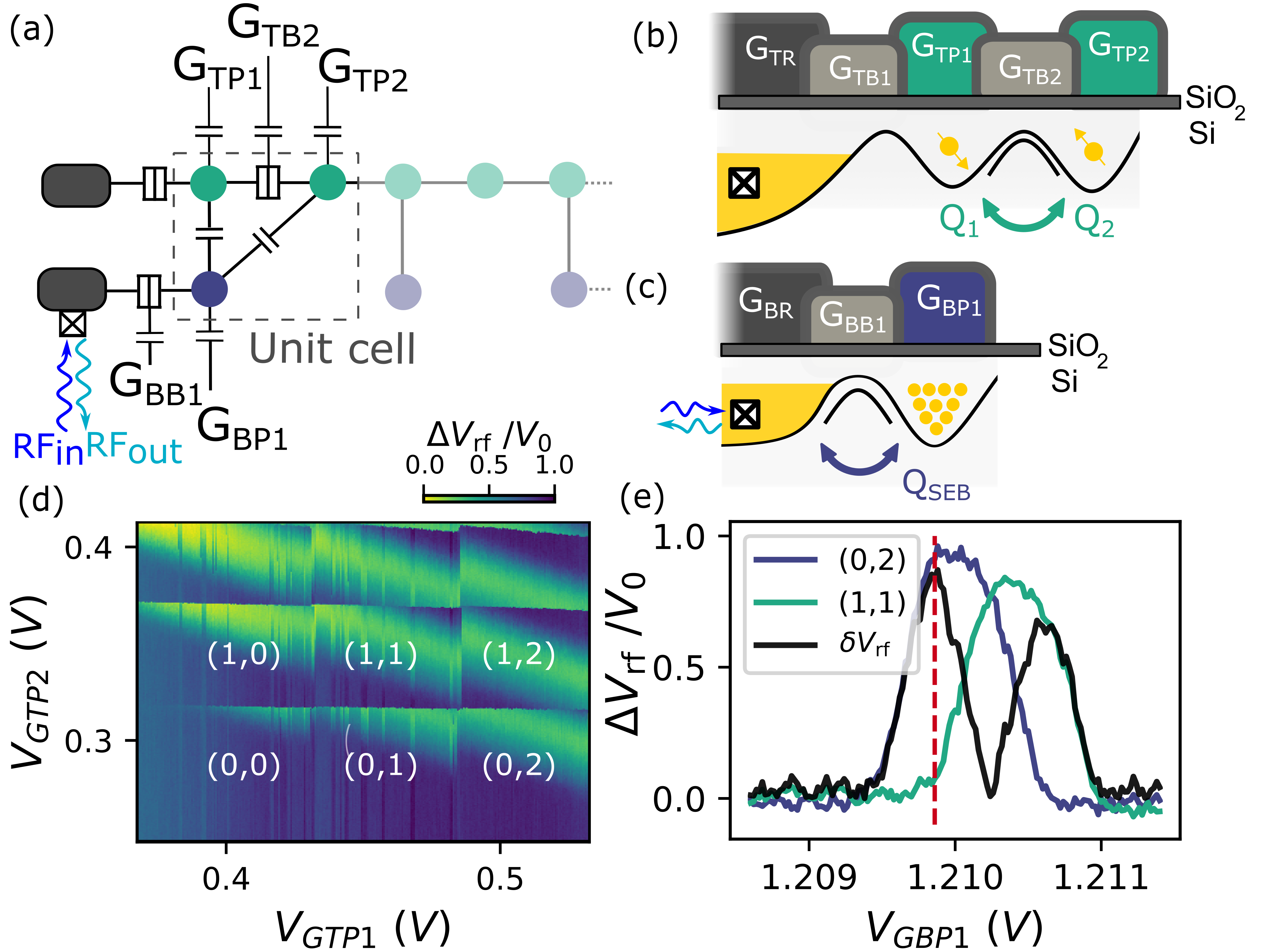}
    \caption{Unit cell for quantum computation in planar silicon MOS with an array of sensors and qubits. (a) Schematic of the device showcasing a scalable architecture for silicon QPU with a repeating unit cell containing one sensor dot (purple) and two-qubit dots (blue). QDs are represented by a circle, and tunnel barriers by a capacitance in parallel to a resistance. The dots and barriers are capacitively coupled to gates. (b) 
    Cross-sectional view of the top row, tuned as a DQD with $Q_1$ and $Q_2$ charges, respectively, under plunger gates $G_{TP1}$ and $G_{TP2}$. The barrier gate TB2 regulates the tunnel coupling between dots. (c) Cross-section view of the bottom row with the sensor SEB tuned in the many-electron regime. An RF tone is sent to the ohmic contact of that reservoir gate for reflectometry. The barrier gate BB1 regulates the tunnel coupling between the dot and reservoir. (d) Charge stability diagram of the qubit-dots with electron occupation down to the first electrons. (e) Coulomb peak of the SEB when the qubit array is in the (1,1) or (0,2) electron occupation. We maximise the difference in signal $\delta V_{\rm rf} = |\Delta V_{\rm rf}^{(11)} - \Delta V_{\rm rf}^{(02)}|/V_0$ by operating at the red dashed line.}
\label{fig:fig1}
\end{figure}

We consider a QPU architecture consisting of a bilinear QD array, 
shown in Fig.~\ref{fig:fig1}(a). This design, where the sensor is not co-linear with the qubits, can be scaled laterally without a reduction in sensitivity and is compatible with the requirements of error-correcting codes \cite{siegel2024towards}. The device studied here can be seen as a unit cell within the larger array of Fig.~\ref{fig:fig1}(a) and is fabricated using an industrial-grade 300~mm wafer planar MOS $^\text{nat}$Si process~\cite{Camenzind2021}. The top row features a DQD with a tunable barrier gate, while the bottom row contains the SEB dot, which is tunnel coupled to an electron reservoir extended from an ohmic implanted region using an accumulation gate (see Fig.~\ref{fig:fig1}(b) and (c)). The ohmic contact is connected to a superconducting LC resonator for readout (see Methods).

Charge sensing is performed using the SEB dot in the many-electron regime, monitoring a Coulomb blockade peak produced by an electron cyclically tunneling between the dot and the reservoir. This enables sensing of the top row DQD in the single-electron regime, as evidenced by the charge stability diagram in Fig.~\ref{fig:fig1}(d). To perform Pauli Spin Blockade (PSB) we focus on the region of the interdot charge transition (ICT) between the (1,1) and (0,2) charge states of the DQD and adjust the SEB plunger gate BP1 to the point of maximal contrast as highlighted in Fig.~\ref{fig:fig1}(e). 

We perform PSB on the ICT shown in Fig.~\ref{fig:fig2}(a) and apply a static in-plane magnetic field ($B_0 =$ 100~mT) which lifts the degeneracy between triplet states, resulting in the energy levels of Fig.~\ref{fig:fig2}(b). We perform a pulse sequence as laid onto Fig.~\ref{fig:fig2}(a-b) to prepare a superposition of the three possible two-electron states $\ket{\rm S}= \frac{1}{\sqrt{2}}\left( \ket{\uparrow\downarrow} - \ket{\downarrow \uparrow} \right)$, $\ket{\rm T_0}=\frac{1}{\sqrt{2}}\left( \ket{\uparrow\downarrow} + \ket{\downarrow \uparrow} \right)$ or $\ket{\rm T_-}=\ket{\downarrow\downarrow}$ (see Methods). 
At the measurement point `M' in the (0,2) region, the singlet state $\ket{\rm S}$ rapidly transitions from (1,1) to (0,2) whereas triplets $\ket{\rm T_-}$ and $\ket{\rm T_0}$ are blocked in (1,1) until they relax to the singlet (0,2) ground state at the respective rates $\Gamma_{\rm T_-}$ and $\Gamma_{\rm T_0}$.
The blockade can already be evidenced in the gate voltage stability diagram by the persistence of the (1,1) signal in the vicinity of the ICT (inset in Fig.~\ref{fig:fig2}(a)). PSB can be lifted when the measurement point becomes sufficiently deep in detuning such that an excited triplet (0,2) state is accessible. From the width of the readout window and the estimated ICT lever arm of $\alpha_\text{ICT} = 0.17$ (see Supplementary Note~\ref{app: leverarm}), we extract an excited state energy in dot TP2 of 25~$\upmu$eV, likely arising from a low-lying excited valley state~\cite{yang2013spin, spence2022spin}.

PSB can be further studied by monitoring the signal in the time domain immediately after pulsing to the measurement point, with three typical single-shot traces shown in Fig.~\ref{fig:fig2}(c). 
Signals corresponding to (0,2) configuration at the measurement point can be ascribed to $\ket{\rm S}$ while other traces that show the (1,1) configuration for a finite time before relaxation, are assigned to one of the two triplets $\ket{\rm T_-}$ or $\ket{\rm T_0}$, distinguishable from their relaxation rates $\Gamma_{\rm T_-}\ll\Gamma_{\rm T_0}$ (see Methods). 

This work considers two approaches to classify the qubit state of a single-shot trace. The first, and commonly used method (referred to as the \emph{threshold method}) is to average the time-domain signal over some window $t_{\rm read}$ and compare the average to a predefined threshold. The second approach uses a Gaussian Hidden Markov Model (HMM)~\cite{murphy2012machine, martinez2020improving} to infer the state by making use of all data points within the time window ($t_{\rm read}$) of the time trace (see Methods). The threshold method distinguishes between two states based on the integration time $t_{\rm read}$ and the relevant relaxation times~\cite{Seedhouse2020}. Setting $t_{\rm read}\ll\Gamma_{\rm T_0}^{-1}$ enables \textit{singlet-triplet} readout (discriminating $\ket{\rm S}$ from ($\ket{\rm T_-}$,$\ket{\rm T_0}$), while setting $\Gamma_{\rm T_0}^{-1}\ll t_{\rm read}\ll\Gamma_{\rm T_-}^{-1}$ achieves \textit{spin parity} readout --- discriminating even-parity ($\ket{\rm T_-}$) from odd-parity ($\ket{\rm S}$, $\ket{\rm \rm T_0}$) states. As we shall see below, thanks to the difference in relaxation rates, the HMM approach can go beyond readout in the two basis and use a single-shot trace to distinguish three sets of states: $\ket{\rm S}$, $\ket{\rm \rm T_0}$ and $\ket{\rm T_-}$.

\section{Results}
\subsection{Accurate modelling of three-state distribution}
In calculating qubit readout fidelity, a common challenge concerns discerning the effects between State Preparation And Measurement (SPAM) errors, often measured by comprehensive tomography techniques~\cite{knill2008randomized, mills2022high} or repetitive measurements~\cite{Huang2023}. Alternatively, to estimate the contribution due solely to readout errors, it is possible to model the distribution of measurement results using an analytical or computational model based on underlying assumptions on the physical system. In this case, the reported readout fidelity is highly dependent on the validity of the model, as well as the quality of the fit to the experimental data~\cite{struck2021robust}.

\begin{figure}
    \centering
    \includegraphics[width=0.95\linewidth]{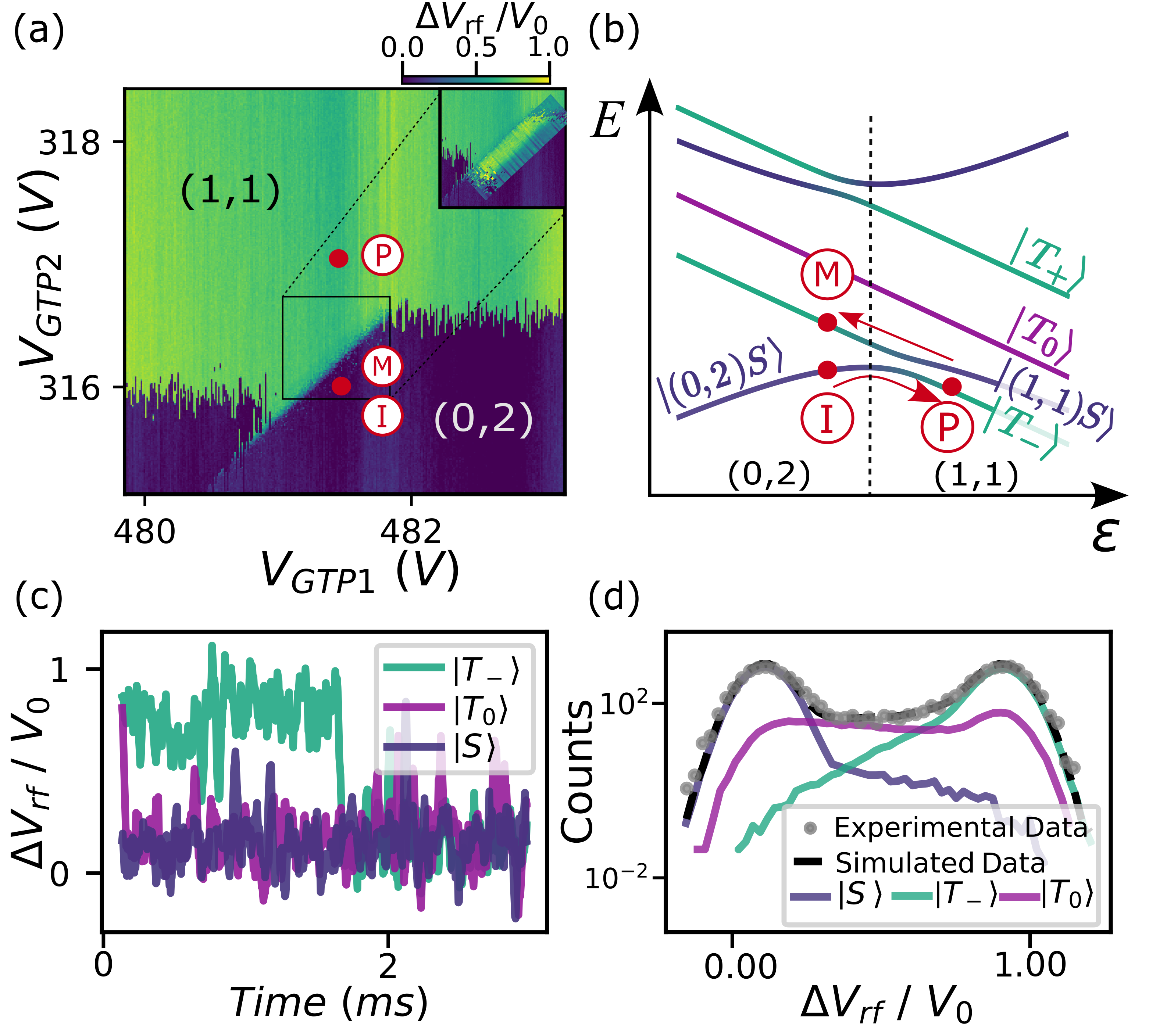}
        \caption{Spin readout protocol. (a) Charge stability diagram of the qubit array around the (1,1)-(0,2) ICT, with the readout voltage pulse sequence overlaid. The sequence consists of three steps: ‘I’ for initialisation to \(\ket{ \rm S(0,2)}\), ‘P’ for plunge to (1,1), and ‘M’ for measurement. The sequence is detailed in Methods. The PSB window with the persistence of (1,1) signal is shown in the inset. (b) Energy levels for a DQD with two electrons, as a function of detuning ($\varepsilon$), the potential difference between the two QDs. At a finite magnetic field, the Zeeman energy splits the triplet states, and the singlet is coupled to the polarised triplet via spin-orbit coupling. The pulse sequence points detailed in the Methods are also overlaid.  (c) Single-shot traces for $\ket{ \rm S}$,  $\ket{ \rm T_-}$ , and $\ket{ \rm T_0}$. (d) Histogram of 10,000 single-shot traces averaged over the first $t_{\rm read} =204\ \upmu s$ after correcting for sensor drift (see Supplementary Note~\ref{app: pre-processing}) (scatter plot), alongside histograms of $\ket{ \rm S}$,  $\ket{ \rm T_-}$, $\ket{ \rm T_0}$ and their sum (black dashed line) for data simulated via the HMM (see Supplementary Note~\ref{app: data_simulation}). 
        The dataset reveals decay times $\Gamma^{-1}_{\rm T_-} = 290$ ms  and $\Gamma^{-1}_{\rm T_0}= 0.170$ ms. The simulated data includes the effect of a two-level fluctuator for a better match with the experimental data.}
\label{fig:fig2}
\end{figure}

We begin by assessing readout fidelity using the threshold method and consider the expected distribution of the averaged value of a single-shot trace. Figure~\ref{fig:fig2}(d) shows the histogram of 10,000 time-average single-shot traces (after compensating for sensor drift, see Supplementary Note~\ref{app: pre-processing}). A common approach to interpret such data is to consider a bimodal probability density model~\cite{Barthel2009}, assuming that only two spin states are involved: $\ket{ \rm S}$ and one of the triplet states such as $\ket{ \rm T_-}$ if parity readout is performed. The readout fidelity and optimal threshold are calculated from this modelled distribution (see Methods). In the case of Gaussian noise, the readout fidelity can be related to the voltage SNR of the time-dependent spin readout signal~\cite{Gambetta2007}:
 \begin{equation}
 \label{eq:genfid}
    F_\text{m}^*=\frac{1}{2}
    \left[1+\text{erf}\left(\frac{{\rm {SNR}}(t_{\rm read})}{2\sqrt{2}}\right)
    \exp\left(-\frac{\Gamma t_{\rm read}}{2}\right)\right],
 \end{equation}
 where, $\Gamma$ is the relaxation rate of the relevant triplet.

A clear limitation of this two-state model is its failure to account for the additional states of the two-spin system, and this is particularly apparent when $t_{\rm read}$ is similar to $\Gamma_{\rm T_0}^{-1}$ and much smaller than $\Gamma_{\rm T_-}^{-1}$.  
In this case, the expected contribution from the $\ket{\rm T_0}$ state is evident at the intersection of the peaks from $\ket{\rm S}$ and $\ket{\rm T_-}$  (see also Fig.~\ref{fig:histogram_barthel} in Methods) and is not accounted by the two-state model. This introduces additional measurement errors, such that readout fidelity is overestimated, as we shall see. To improve the model accuracy, we simulate the probability distribution using single-shot traces generated by a HMM which includes three hidden states ($\ket{ \rm S}$, $\ket{ \rm T_0}$, and $\ket{ \rm T_-}$) as well as noise from a two-level fluctuator (TLF) (see Methods). This allows an accurate modelling of the system, as seen in the simulated histogram of Fig.~\ref{fig:fig2}(d) which matches well with the experimental data. Below, we compare measures of parity readout fidelity using either the two-state model or the HMM for data simulation, in both cases distinguishing odd/even parity states with the threshold method. This highlights the regime where two-state modelling is insufficient to accurately estimate readout fidelity. We use $F_\text{m}^*$ to refer to the (typically overestimated) value using the simple two-state model and $F_\text{m}$ using the more comprehensive HMM. 

\subsection{Maximising charge readout signal}

Regardless of the method used to model single-shot data, improving the charge sensor SNR provides a route to enhance spin readout fidelity and speed, as evidenced by Eq.~\ref{eq:genfid}. 
For an SEB, the signal amplitude (and hence the SNR) is related to the change in capacitance $\Delta C_{\rm DRT}$ between blocked and fully degenerate charge states of the SEB due to the dot to reservoir transition (DRT) (see Supplementary Note~\ref{app: RF opti}). Experimentally, the maximum signal is the amplitude of the Coulomb peak  $\Delta V_{\rm rf}^{\rm max}$, however for charge sensing the relevant signal derives from the fractional change in capacitance caused by a charge sensing event --- e.g. transitions between (1,1) and (0,2) in the DQD --- and is given by the contrast $\delta V_{\rm rf}$. These two quantities are related by $ \delta V_{\rm rf} = \eta  \Delta V_{\rm rf}^{\rm max}$ with $\eta$ a dimensionless parameter bounded between 0 and 1 such that $\eta=$1 when the (0,2) and (1,1) peak separation is much greater than the SEB Coulomb peak linewidth and $\eta=$0 when they fully overlap. For charge readout, we aim at maximising the product of $\eta$ and $\Delta C_{\rm DRT}$, with the latter following the expression~\cite{Peri2023}: 
\begin{equation}
\label{eq:cap_seb}
    \Delta C_{\rm DRT} \simeq \frac{4(1-\alpha_\text{DRT})^2 e^2}{3k_BT_e} \underbrace{\frac{1}{1+(f_{RF}/\gamma)^2}}_{(i)} \underbrace{\frac{1}{1+h\gamma/k_B T_e}}_{(ii)},
\end{equation} 
where $\alpha_\text{DRT}$ is the DRT lever arm, $e$ the electron charge, $k_B$ Boltzmann's constant, $h$ Planck's constant, $T_e$ the electron temperature, $f_\text{RF}$ is the RF frequency and $\gamma$ is the DRT tunnel rate which can be controlled by barrier gate voltage. 

Equation~\ref{eq:cap_seb} reveals two competing terms in the change in capacitance as $\gamma$ increases: (i) is monotonically increasing as it gets more likely for the electron to tunnel during a RF cycle
(ii) is monotonically decreasing as the charge transition becomes lifetime broadened, reducing the probability of tunnelling. A sweet spot is found when $\gamma$ is greater than $f_\text{RF}$ --- here 570~MHz --- but still lower than $k_BT_e$ where thermal broadening occurs --- 1.87~GHz for our device where $T_e=90$~mK (see Supplementary Note~\ref{app: leverarm}). 

\begin{figure}
    \centering
    \includegraphics[width=\linewidth]{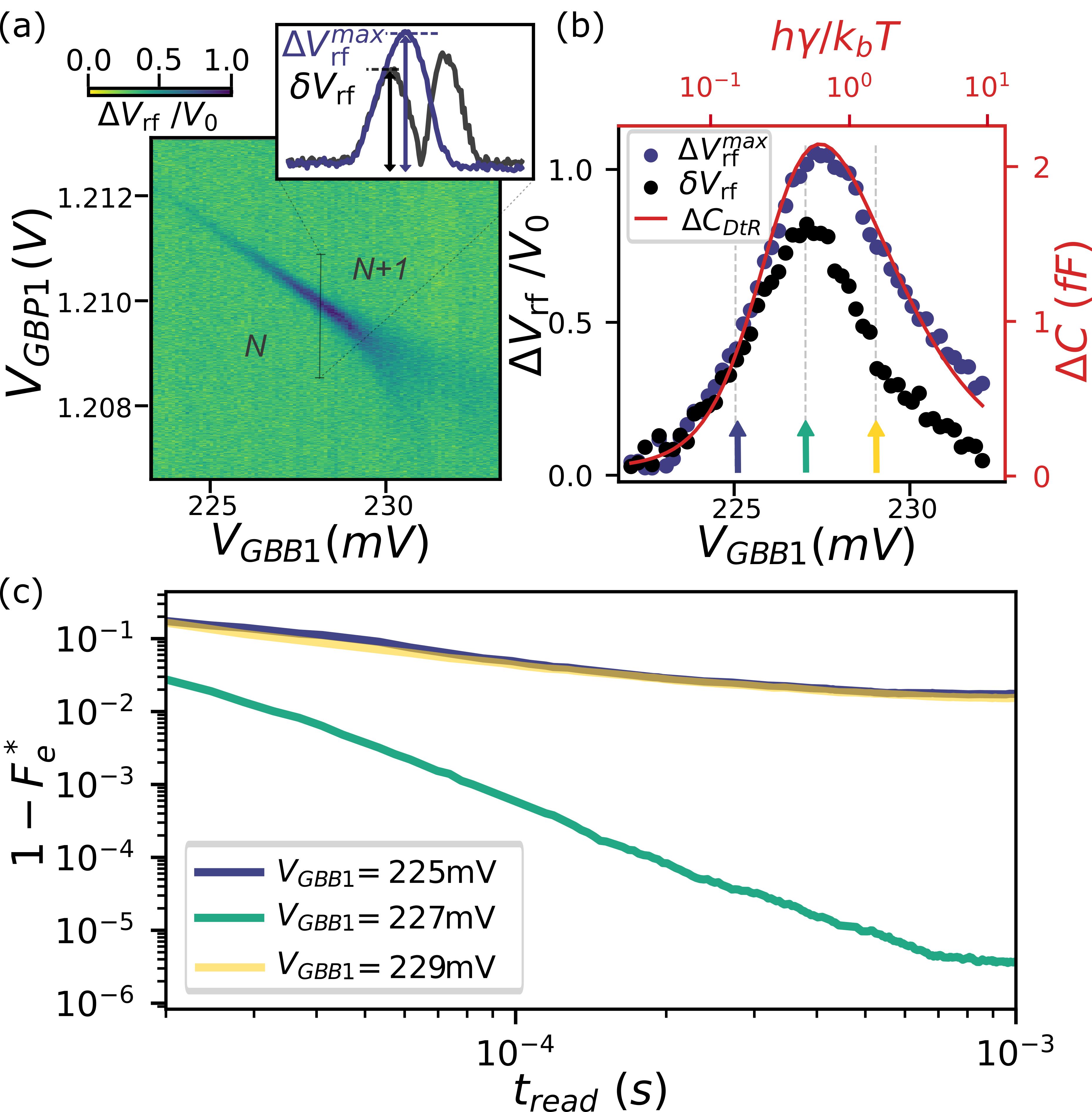}
    \caption{Tuning charge readout fidelity via SEB barrier gate. (a) Stability diagram of sensor with the plunger BP1 and barrier gate BB1, showing broadening of the peak at large $V_{\rm GBB1}$ from enhanced lifetime. The peak separates regions of different electron numbers $N$ and $N+1$ where $N\simeq15$ (see Supplementary Note~\ref{app: elecnum}). The inset is the SEB peaks from Fig.~\ref{fig:fig1} (c) where we define $\Delta V_{\rm rf}^{\rm max}$ as the height of the SEB peak (purple) and $\delta V_{\rm rf}$ the contrast between the two charge states (black). (b) Lifetime broadening of the SEB peak, showing the SEB peak height is reduced at $\gamma$ larger than $k_BT$. We superimpose a plot of $\Delta C_{\rm DRT}$ as a function of tunnel rate calculated from Eq.~\ref{eq:cap_seb} with $f_{RF}=576$MHz, lever arm and electron temperature from Supplementary Note~\ref{app: leverarm} and tunnel rate $\gamma$ ranging from 0.05-19 GHz, assuming $V_{\rm GBP1}$ exponentially modulates the tunnel rate $\gamma$. (c) Electrical fidelity for three different SEB barrier gate voltages indicated by arrows in panel (b). The electrical fidelity is calculated 
     using the threshold method and bimodal fitting as detailed in Methods. The optimal point (green) corresponds to $\gamma=1.1$ GHz. The yellow point is in the lifetime-broadened regime, and the purple point is in the slow-tunnelling regime. 
    }
\label{fig:fig3}
\end{figure}

We observe both contributions in Fig.~\ref{fig:fig3}(a), where we monitor the charge sensor signal against the barrier gate BB1. We notice an initial increase in the peak height $\Delta V_{\rm rf}^{\rm max}$ before a reduction from lifetime broadening. This is even more visible in Fig.~\ref{fig:fig3}(b), where we superimpose on the data the calculated change of capacitance $\Delta C_{\rm DRT}$ against the tunnel rate $\gamma$ calculated from Eq.~\ref{eq:cap_seb}, assuming $\gamma$ depends exponentially on the barrier voltage. This shows good agreement between the model and the data and provides quantitative values for the expected SEB tunnel rates around the maximal signal point found at $\gamma = 1.1 $~GHz.

To optimise charge readout, we also examine in Fig.~\ref{fig:fig3}(b) the behaviour of contrast $\delta V_{\rm rf}$, which includes $\eta$ and $\Delta V_{\rm rf}^{\rm max}$. Firstly, the contrast is approximately $80\%$ of the maximum visibility, indicating good coupling between the charge sensor and DQD. Secondly, we note a slight shift in the peak of the $\delta V_{\rm rf}$ compared to the maximum visibility $\Delta V_{\rm rf}^{\rm max}$, arising because $V_\text{GBB1}$ affects not only the amplitude of the peak but also $\eta$ through a modification of the capacitive coupling between the SEB and the DQD. In particular, at larger barrier gate bias, lifetime broadening negatively impacts $\eta$, thereby reducing the contrast $\delta V_{\rm rf}$ further than $\Delta C_{\rm DRT}$.

The impact of SEB barrier gate on readout errors is visible when calculating the electrical fidelity, $F_\text{e}^*$, which imposes an upper bound on the spin readout fidelity and obtained from the two-state model by neglecting any relaxation of the triplet states (i.e.~from Eq.~\ref{eq:genfid} with $\Gamma$ set to 0 - see Methods). It is plotted in Fig.~\ref{fig:fig3}(c) as a function of integration time $t_\text{read}$, at different $V_\text{GBB1}$. We observe an increase in the fidelity with $t_\text{read}$ --- which is expected under Gaussian noise when the SNR is proportional to $\sqrt{t_\text{read}}$ --- until it plateau indicating the onset of $1/f$ charge noise~\cite{von2024multimodule} (see Supplementary Note~\ref{app:snr_chargenoise}). We also notice an improvement in readout at the optimal DRT tunnel rate, which illustrates the importance of tuning the barrier gate of the SEB to maximise the bounds on qubit readout fidelity.

\subsection{Optimising relaxation for parity readout}
\label{sec:level4A}
\begin{figure}
    \centering
    \includegraphics[width=1\linewidth]{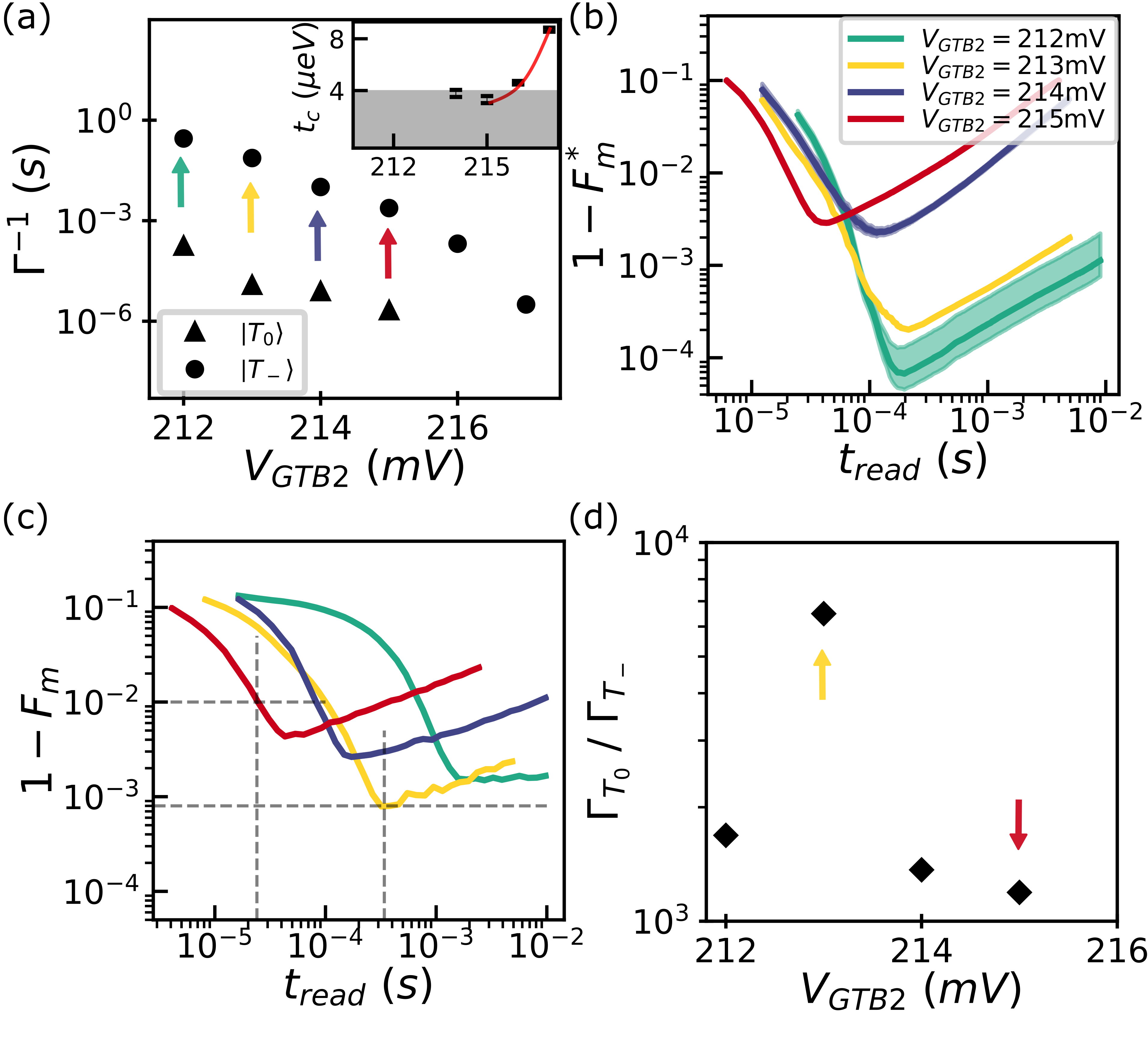}
            \caption{Tuning triplet relaxation time and spin parity readout fidelity using threshold method via qubit barrier gate. (a) $\Gamma^{-1}_{\rm T_-}$ and  $\Gamma^{-1}_{\rm T_0}$ show an exponential dependency on barrier gate voltage. For $V_{\rm GTB2}<$ 212~mV, the dots become too weakly coupled, and the ICT size becomes smaller than the voltage resolution, making the measurement of PSB impossible. For $V_{\rm GTB2}>$ 217~mV, the two QDs begin to merge, and the relaxation times of both triplets are below the measurement bandwidth of 3.3$~\upmu$s. The inset shows the tuning of tunnel coupling with barrier gate voltage, linewidth saturates at 4 $\upmu$eV for $V_{\rm GTB2}<$ 215 mV (grey area) due to thermal broadening with the qubit electron temperature $T_e$= 40 mK (Supplementary Note~\ref{app: tunnelcoupling}). (b-c) Spin parity fidelity as a function of qubit barrier gate voltage. Fidelity is calculated by using the threshold method and simulating the spin distribution using (b) the simple two-state model or (c) the HMM that accounts for a three-state distribution and TLF. In both cases, the initial distribution of states is fixed to $p(\ket{S})=0.25$, $p(\ket{T_0})=0.25$, $p(\ket{T_-})=0.5$ which corresponds to 50/50 odd-even parity distribution and therefore does not bias the overall fidelity towards even or odd parity. We note that the lower barrier gate dataset (red curve) was taken at a different sensor tuning point, hence the slightly different SNR.
            In (c), we highlight with dashed lines the points of highest fidelity (yellow curve --- $F_{\mathrm m}= 99.92\%$ in $340~\upmu$s) and of fastest fidelity (red curve --- $F_{\mathrm m}= 99\%$ in $24\mu$s). 
            (d) Ratio of relaxation times for triplet states $\ket{ \rm T_-}$ and $\ket{ \rm T_0}$. This shows the optimal point for parity readout is at the maximum ratio of relaxation times for $V_{\rm GTB2}=213$ mV (yellow arrow), whereas fast readout is achievable for overall shorter relaxations
            at $V_{\rm GTB2}=215$ mV (red arrow). }
        \label{fig:fig4}
\end{figure}

The simple two-state model of Eq.~\ref{eq:genfid} predicts that spin measurement can be improved by maximising the relevant triplet relaxation times. This can be achieved by adjusting the barrier gate voltage $V_{\rm GTB2}$ as it controls exchange and tunnel coupling between the two dots. We show the latter in the inset of Fig.~\ref{fig:fig4}(a) even though only in a narrow range limited by thermal broadening (Supplementary Note~\ref{app: tunnelcoupling}).
We directly measure the relaxation rates of both $\ket{ \rm T_0}$ and $\ket{ \rm T_-}$ in Fig.~\ref{fig:fig4}(a) and find an exponential dependence on $V_{\rm GTB2}$. As expected, reducing $V_{\rm GTB2}$ and therefore $t_\mathrm{c}$ leads to a strong suppression of triplet relaxation as the dots decouple~\cite{Fogarty2018}. This large tunability of almost three orders of magnitude in triplet relaxation rate enables a study of readout fidelity.

Figure~\ref{fig:fig4}(b) shows the parity readout fidelity $F_{\mathrm m}^*$ predicted from the two-state model for different $V_\text{GTB2}$. Fidelities initially increase as the $t_\text{read}$ lengthens (due to higher SNR) and fall again as relaxation of the measured state introduces errors. We also observe that fidelity is maximised for extended relaxation of triplet states, as expected from Eq.~\ref{eq:genfid}. 
However, such a model is particularly problematic when $\Gamma_{\rm T_0}^{-1}$ is comparable to the integration time, which occurs for $t_{\rm read}$ between 2-200~$\upmu$s given the range of $V_\text{GTB2}$ studied here. In this regime, $\ket{ \rm T_0}$ states, which have odd parity, have not fully decayed, and the presence of this third spin state leads to a significant likelihood of misclassification. The result is an overestimation of the readout fidelity; therefore, requiring a more comprehensive model comprising $\ket{ \rm T_0}$.

This is obtained from using the HMM with three-states and a TLF to model the distribution --- still using the threshold method for classification. An accurate measure of the readout fidelity, $F_{\mathrm m}$, is plotted in Fig.~\ref{fig:fig4}(c). 
Firstly, the maximum fidelity is lower than $F_{\mathrm m}^*$, which assumes a two-state model and so fails to account for (undecayed) $\ket{ \rm T_0}$ states. This is particularly visible at slow relaxation rates and minimal barrier gate voltage, where the long decay of $\ket{ \rm T_0}$ increases the chances of misclassifying it as $\ket{ \rm T_-}$, leading to higher error rates in parity readout. Secondly, accounting for the full distribution changes the tuning point of best fidelity. Although there is a general trend for increasing fidelity with decreasing relaxation rates, due to the ability to increase $t_{\rm read}$ and suppress noise, the optimum fidelity is obtained when maximising the distinguishability of $\ket{ \rm T_0}$ and $\ket{ \rm T_-}$. This occurs when the ratio of relaxation times $\Gamma_{\rm T_0} / \Gamma_{\rm T_-}$ (plotted in in Fig.~\ref{fig:fig4}(d)) is maximised, at which we find a readout parity fidelity of $F_{\mathrm m}= 99.92\%$ in $340~\upmu$s. Some quantum computing applications may prioritise measurement speed over fidelity, as long as the latter exceeds some threshold, and our results show how the relaxation rates can be adjusted to optimise against target parameters. Because we require $\Gamma^{-1}_{T_0} < t_{\rm read} < \Gamma^{-1}_{T_-}$, increasing measurement speed while maintaining high fidelity means increasing the relaxation rates. For example, a fidelity of $F_{\mathrm m}= 99\%$ in $24~\upmu$s can be achieved by increasing $V_{\rm GTB2}$, hence increasing $\Gamma_{\rm T_0}$ and $\Gamma_{\rm T_-}$.

\subsection{Spin classification beyond the threshold method}

\begin{figure*}
    \centering
    \includegraphics[width=0.8\textwidth]{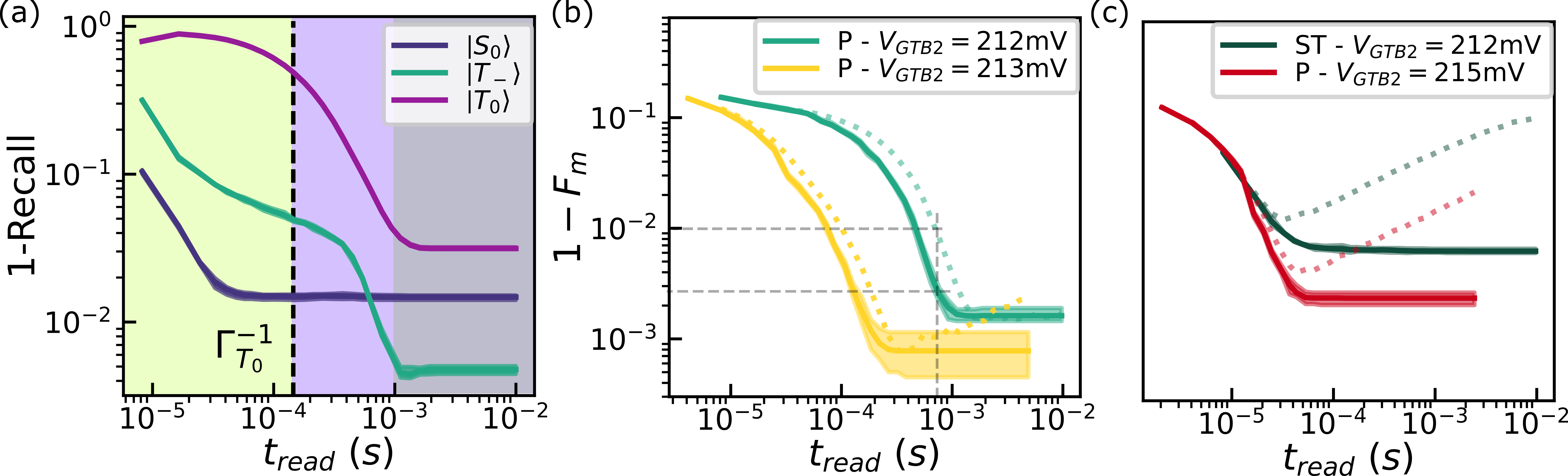}
    \caption{Classification using HMM.
    (a) Recall (proxy to readout fidelity) for each state as a function of the readout time, using HMM to model the data and to classify traces at $V_{\rm GTB2} = 212$~mV. For $t_\mathrm{int}<<\Gamma_{ \rm T_0}^{-1}$ (light green region), measurement of $\ket{ \rm T_-}$ and $\ket{ \rm S}$ improves with readout time due to noise reduction, while un-relaxed $\ket{ \rm T_0}$ is misidentified as $\ket{ \rm T_-}$. For $t_\mathrm{read} \geq \Gamma_{ \rm T_0}^{-1}$ (purple region), $\ket{ \rm T_0}$ can be distinguished from $\ket{ \rm T_-}$, leading to an improvement in the readout of both triplets. Finally, at $t_\mathrm{read}>>\Gamma_{ \rm T_0}^{-1}$ (grey region), all ${\ket{ \rm T_0}}$ have decayed thereby no additional information can be gained which causes the recall to plateau for all states.  (b-c) Readout fidelity calculated using HMM to model the data and to classify traces ($F_{\rm m}$ in thick line), compared to using threshold to classify traces ($F_{\rm m}^*$ in dashed line). Readout is done in the parity (P) or singlet-triplet (ST) basis. Panel (b) highlights the speed improvement of HMM over threshold in the case of long $\Gamma_{ \rm T_0}^{-1}$ for readout times comparable to that relaxation rate. The optimal case shows a two-fold speed improvement for $V_{\rm GTB2} = 212$~mV from 99\% to 99.5\% at the same readout time, as indicated by the grey dashed line. Panel (c) highlights the overall fidelity improvement as relaxation gets faster and therefore SNR can be further increased. We note that readout in the ST basis at 212 mV ($\Gamma_{ \rm T_0}^{-1} = 170~\upmu$s is the limiting relaxation) should be equivalent to parity readout at 216 mV where $\ket{T_0}$ should be all decayed before the smallest readout time and $\Gamma_{ \rm T_-}^{-1}\simeq200~\upmu$s from Fig.~\ref{fig:fig4}(a).
    }
    \label{fig:fig5}
\end{figure*}

Finally, we explore whether it is possible to improve readout spin and fidelity by going beyond the threshold method and using the full information available from a single shot trace~\cite{Gambetta2007, d2014optimal, d2016maximal, martinez2020improving}. We use the forwards-backwards algorithm based on the HMM introduced before to classify a single-shot trace without averaging (see Methods).
Taking advantage of the different relaxation rates of the $\ket{ \rm T_0}$ and $\ket{ \rm T_-}$ states, this approach discriminates all three states which we prepare here --- $\ket{ \rm S}, \ket{ \rm T_0}$ or $\ket{ \rm T_-}$. For the classification to be accurate, it is necessary to wait for the decay of all the states but one in order to produce a characteristic signature of the time trace 
This effect on readout accuracy is visible when measuring the \emph{recall} as a function of readout time. Recall is defined as the ratio of correctly identified traces for a given state to the total number of occurrences of that state, and serves as the best proxy for the measurement fidelity $F_{\mathrm m}$ of a single state (see Methods). In Fig.~\ref{fig:fig5}(a), we calculate the recall of each state in the regime of longest triplet relaxation rates, where the effect of undecayed $\ket{T_0}$ states is most visible. For readout times much smaller than $\Gamma_{T_0}^{-1}$, it is hard to accurately classify $\ket{T_0}$ states, as there is no way to distinguish them from $\ket{T_-}$. The classification becomes possible only after the odd-parity triplets have started to decay (i.e. for $t_{\rm read} >\Gamma_{ \rm T_0}^{-1}$), which further benefits the readout of $\ket{T_-}$. Moreover, fidelity no longer reduces at long readout times as the HMM effectively operates as a non-linear Bayesian filter \cite{Gambetta2007}. Such an analysis highlights the key difference between classifying single-shot traces with the threshold method and the forward-backwards algorithm. It informs us on how long it is best to measure for, depending on which state is most relevant for computation. 

Having classified a single-shot trace into these three sets, we can combine the results in different ways to obtain a binary readout in the singlet-triplet or parity bases. In some cases, this offers an improvement in fidelity and speed over the threshold method, as shown in Fig.~\ref{fig:fig5}(b-c). Speed improvement is obtained in the case of slow $\ket{T_0}$ relaxation, where the binary threshold classification struggles to distinguish between three states. The forwards-backwards algorithm enables up to a two-fold reduction in error rate for the same readout time, as shown in panel (b). In the case of fast relaxation, the algorithm increases the overall fidelity as it is no longer limited by the decay at long relaxation times, as shown in panel (c).

\section{\label{sec:level5} Discussion}

\begin{figure}
    \centering
    \includegraphics[width=0.95\linewidth]{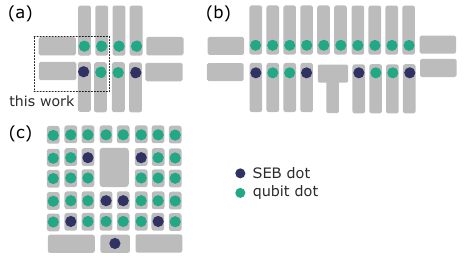}
    \caption{QPU architectures containing the SEB. (a) 2$\times$4 array of QD architecture based on the unit cell presented in this work (dashed square). Notice that in the device studied in this work, the qubit dot positioned in the same row as the sensor dot was non-functional. (b) 2$\times$N array with T shape reservoir where sensor and qubits can be placed on the lower row. The optimal number of qubit dots between sensors in the bottom row remains to be determined and will depend on the quality of the sensor. (c) N$\times$N array with island reservoirs leading to higher quit connectivity. The size of the reservoir gate would also have to be optimised: a small reservoir could lead to the formation of elongated dots \cite{patomaki2024elongated} whereas increasing the reservoir size means losing connectivity.}
\label{fig:fig6}
\end{figure}

The compact nature of SEB we have studied here allows it to be incorporated within tileable unit cells in scaled QPU architectures, as illustrated in Fig.~\ref{fig:fig6}.
A key advantage here over previous demonstrations in MOS QDs is that the sensors are no longer confined to the edge of the spin qubit array~\cite{steinacker2024300}. 
In principle, SETs could be used in the second, parallel channel, as has been shown in SiGe~\cite{lawrie2020quantum}, however, the requirement for two SET reservoirs and their screen effects, limits the achievable qubit connectivity.
Beyond 2$\times$N arrays, SEB sensors could be deployed in the middle of an N$\times$N array of dots such as those fabricated by the single-layer etched-defined electrodes (SLEDGE) process \cite{ha2021flexible}. This would require forming island reservoirs, which could be challenging, but possibly achieved by localised doping or accumulation gates with charges implanted via initial illumination of the sample.

The HMM, which enables the forward-backwards algorithm, opens pathways for a more precise readout method. It is particularly effective in cases where readout involves more than two states that produce the same characteristic signal but differ in durations (i.e. different decay rates). In such a scenario, traditional thresholding is limited to a binary output and struggles to distinguish states unless they have decayed for long enough that it affects the averaged signal, leading to a deterioration in measurement speed. Our algorithm could be extended to classifying the $\ket{ \rm T_+}$ state --- as long as it has a different relaxation rate from the other triplets --- and could offer more efficient state initialisation, effectively reducing the number of readout attempts~\cite{nurizzo2023complete, Philips2022}. Moreover, this method may enable on-the-fly measurement with adaptive integration times~\cite{d2016maximal} where the readout stops once it is confident enough of the predicted qubit state, providing faster integration. Finally, going beyond spin qubits, a readout protocol capable of discriminating between multiple states could be particularly useful when measurable states are beyond those in the computational subspace such as leakage states \cite{andrews2019quantifying, ghosh2013understanding, brown2018comparing}. In summary, this work has highlighted the importance of the SEB in building a scalable QPU 2D architecture, and opens the way for scalable and high-fidelity readout of spin qubits.

\section{\label{sec:Method}Methods}

\subsection{Experimental set-up and resonator characterisation}
\label{app: set-up}
The device used is a 300 mm wafer-scale MOS planar structure fabricated at the industrial-grade IMEC facility~\cite{Camenzind2021}. Three layers of overlapping PolySi gates are defined above an unpatterned $^{\text{nat}}$Si
substrate, providing the electrostatic potential to form QDs. Electrons are loaded from a nearby 2D-electron gas extended via accumulation gates from localised reservoirs of dopants. The chip is placed on a Printed Circuit Board (PCB) made of Rogers 4003C material and gold finish, which hosts a tank circuit for reflectometry. 

The LC resonator circuit used for readout comprises a NbTiN spiral inductor $L$ = 88.7~nH and a surface mount coupling capacitor $C_c$ = 0.2~pF. An RF signal is sent to the device after 30~dB of attenuation at different stages of the dilution refrigerator, passband RF filtering with Mini-Circuits VLF-1800+ VLF145+ and a directional coupler Mini-Circuit ZFDC-20-33-S+ at the cold plate stage, adding 12~dB attenuation. The reflected signal is amplified at the 4K stage using a LNF-LNC0.2-3B ultra-low noise cryogenic amplifier and at room temperature with two Mini-Circuit ZX60-112LN. The device is operated in a Proteox MX Oxford Instruments dilution refrigerator at base temperature $T_{\text{MXC}} = 10$mK.

\begin{figure}
    \centering
    \includegraphics[width=0.4\textwidth]{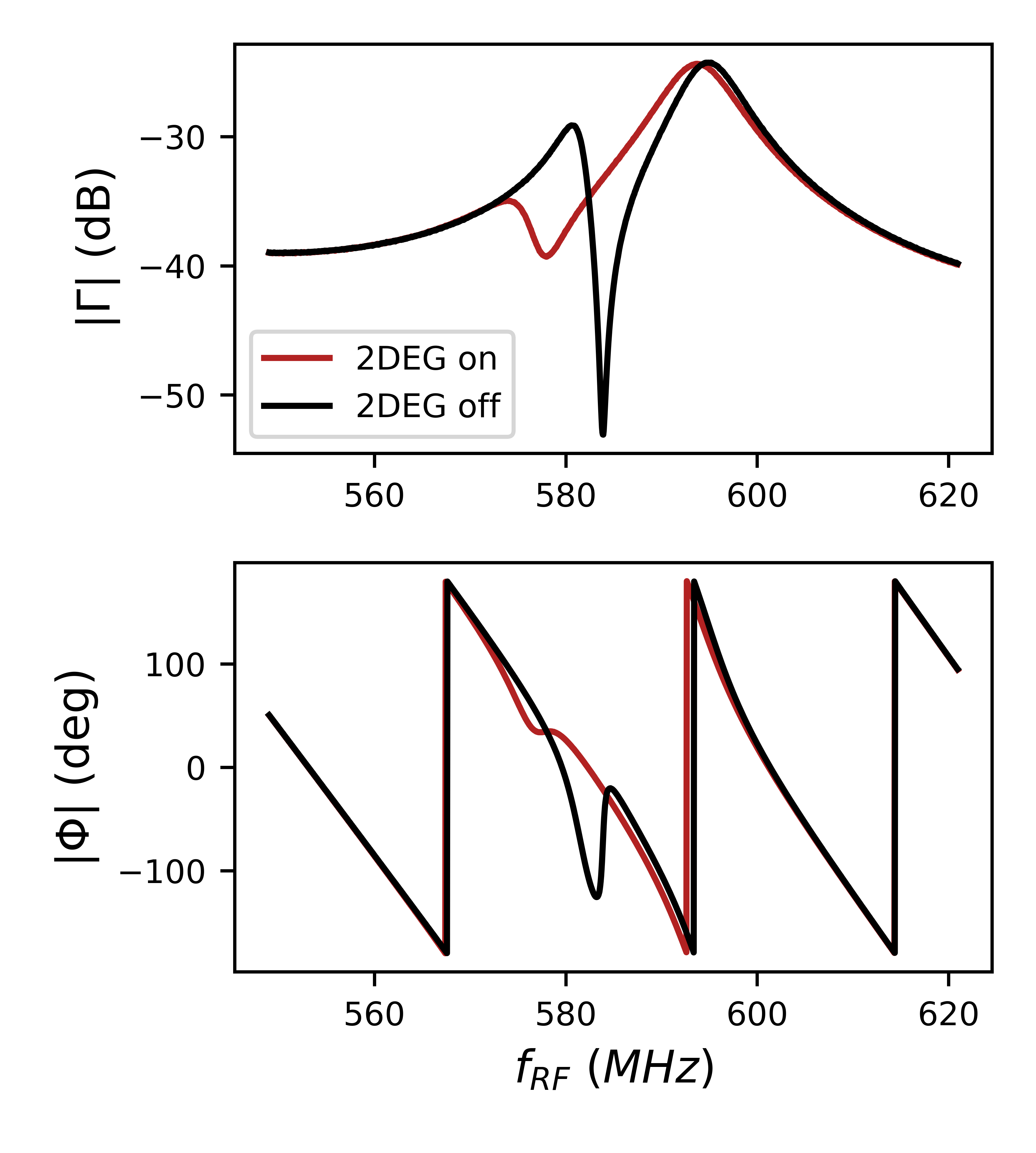}
    \caption{Readout resonator characterisation. VNA traces at 10mK, when bottom reservoir gate is at to turn-on voltage point and electron 2DEG is accumulated (red line) and when all gate voltages are at zero (black line).}
    \label{fig:resonance1}
\end{figure}

To characterise the resonator, the reflected signal $\Gamma$ is analysed by a Vector Network Analyser as shown in magnitude and phase in Fig.~\ref{fig:resonance1}. We measure $\Gamma$  as all the device gates are at zero voltage (black line) and when the 2DEG is extended under the SEB reservoir gate (red line). The shift of the resonance in phase and magnitude is due to changes in the capacitance and resistance from the formation of the 2DEG. After removing the background from fitting the standing wave, we find the resonant frequency $f_0$, the full-width half maximum (FWHM) $\Delta f$ and extract the total quality factor $Q_r = f_0/\Delta f$, the coupling $\beta = \frac{\Gamma_V(f_0)-1}{\Gamma_V(f_0)+1}$, the internal quality factor $Q_{int} = (1+\beta)Q_r$,  the parasitic capacitance $C_p=1/(4\pi^2 f_0^2 L)- C_c$ and the equivalent resistance to ground $R_C = Q_{int} \sqrt{\frac{L}{C_c +C_p}}$. These terms are gathered in Table~\ref{tab:resonator} for both 2DEG off and on. 

\begin{table}[]
    \centering
    \begin{tabular}{c|c|c}
            & 2DEG off & 2DEG on \\
        \hline
         $f_0$ (MHz) &  583.9 & 578.6\\
         $Q_{r}$& 104 & 74.8 \\
         $Q_{int}$& 197 & 106 \\
         $\beta$ & 0.90 & 0.42 \\
         $C_p$ (pF) & 0.55 & 0.56 \\
         $R_C \ (k\Omega)$ & 72.1 & 38.6
    \end{tabular}
    \caption{Parameters of the resonator.}
    \label{tab:resonator}
\end{table}

For charge sensing, the SEB dot is operated in the many-electron regime (see Supplementary Note~\ref{app: elecnum}), with its ohmic contact connected to the LC resonator. When the electrostatic potential of the SEB dot and the reservoir are near alignment, the RF signal induces an electron to cyclically tunnel between them, contributing to the measured electrical impedance of the SEB. Changes in this impedance, caused, for example, by changes in the local potential of the SEB, are detected in the reflected RF signal which is demodulated with a DC1670A I/Q demodulator, low-pass filtered by SR560 Stanford voltage pre-amplifiers with a cut-off frequency of either 10~kHz or 1~MHz, and acquired with an M4i digitiser. The change in signal in the (I, Q) plane is then projected onto the axis that maximises the response and renormalised as $\Delta V_{\rm rf} / V_0$ where $V_0$ is the maximum voltage change.
In the vicinity of a Coulomb blockade peak, the SEB is highly sensitive to its electrostatic environment. Due to their mutual capacitive coupling, changes in the charge configuration of the DQD can be detected by shifts in the SEB signal.

RF frequency and power are optimised to enhance the signal contrast $\delta V_{\rm RF}$ - as introduced in the main text. The RF frequency is optimised at $f_0= 576$ MHz and power to $P_{\rm in} = -80~$dB to the device.

\subsection{Spin preparation and readout}
\label{app: ST-} 

To prepare the appropriate mixture of singlet and triplet states for PSB, we exploit the $S/T-$ anticrossing and the $S-T_0$ dephasing in the (1,1) region. We first wait in the (0,2) region [point `I' in Fig.~\ref{fig:fig2}(a-b)] for 10 times the spin relaxation time so that the system is relaxed in the $\ket{(0,2)S}$ state. 

\begin{figure}
    \centering
    \includegraphics[width=0.75\linewidth]{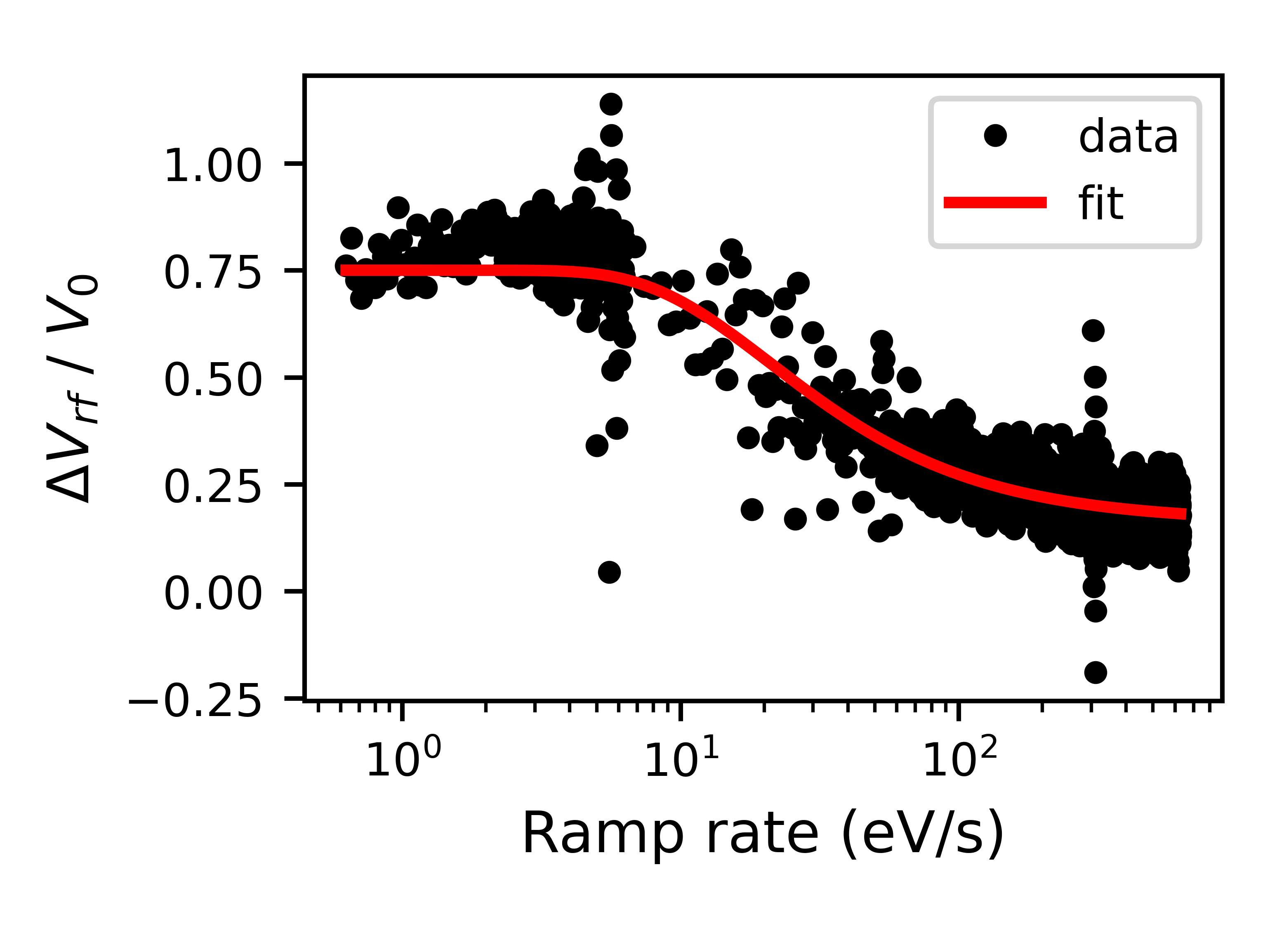}
    \caption{$\Delta_{S/T-}$ characterisation. LZ single passage when varying ramp rate across the anticrossing. Fit to equation \ref{eq:LZsinglepassage} with $\Delta_{S/T_-} = 46.9 \pm 0.9$ neV.}
    \label{fig:LZ single passage}
\end{figure}

Next, we pulse into the (1,1) region [point `P']. Depending on the ramp rate used to traverse the $S/T_-$ avoided crossing, either the $\ket{ \rm S(1,1)}$ or $\ket{ \rm T_-(1,1)}$ state will be populated. To calibrate the avoided crossing, we vary the ramp rate and measure the probability of initialising $\ket{ \rm T_-(1,1)}$, as shown in Fig.~\ref{fig:LZ single passage}. On this plot, we have converted the gate voltage ramp rate to eV/s using lever arm $\alpha = 0.14$ from Supplementary Note~\ref{app: leverarm} and the y-axis is normalised such that high voltage ($\Delta V_{\rm rf}/V_0 =1 $) corresponds to blockaded signal i.e. triplets, and low voltage ($\Delta V_{\rm rf}/V_0 =0$) corresponds to singlets. The step shape of the curve is expected from the probability of a LZ transition given by 
\begin{equation}
\label{eq:LZsinglepassage}
    P_{LZ} = \text{exp}\left(\frac{-2\pi \Delta_{S/T_-}^2}{\hbar v_{in}}\right),
\end{equation}
which we fit to the data to extract $\Delta_{S/T_-} = 46.9 \pm 0.9$ neV. For state preparation, we fix the ramp rate such that it ensures approximately 50\% of the population reaches $\ket{(1,1)T_-}$. 

\begin{figure}
    \centering
    \includegraphics[width=0.8\linewidth]{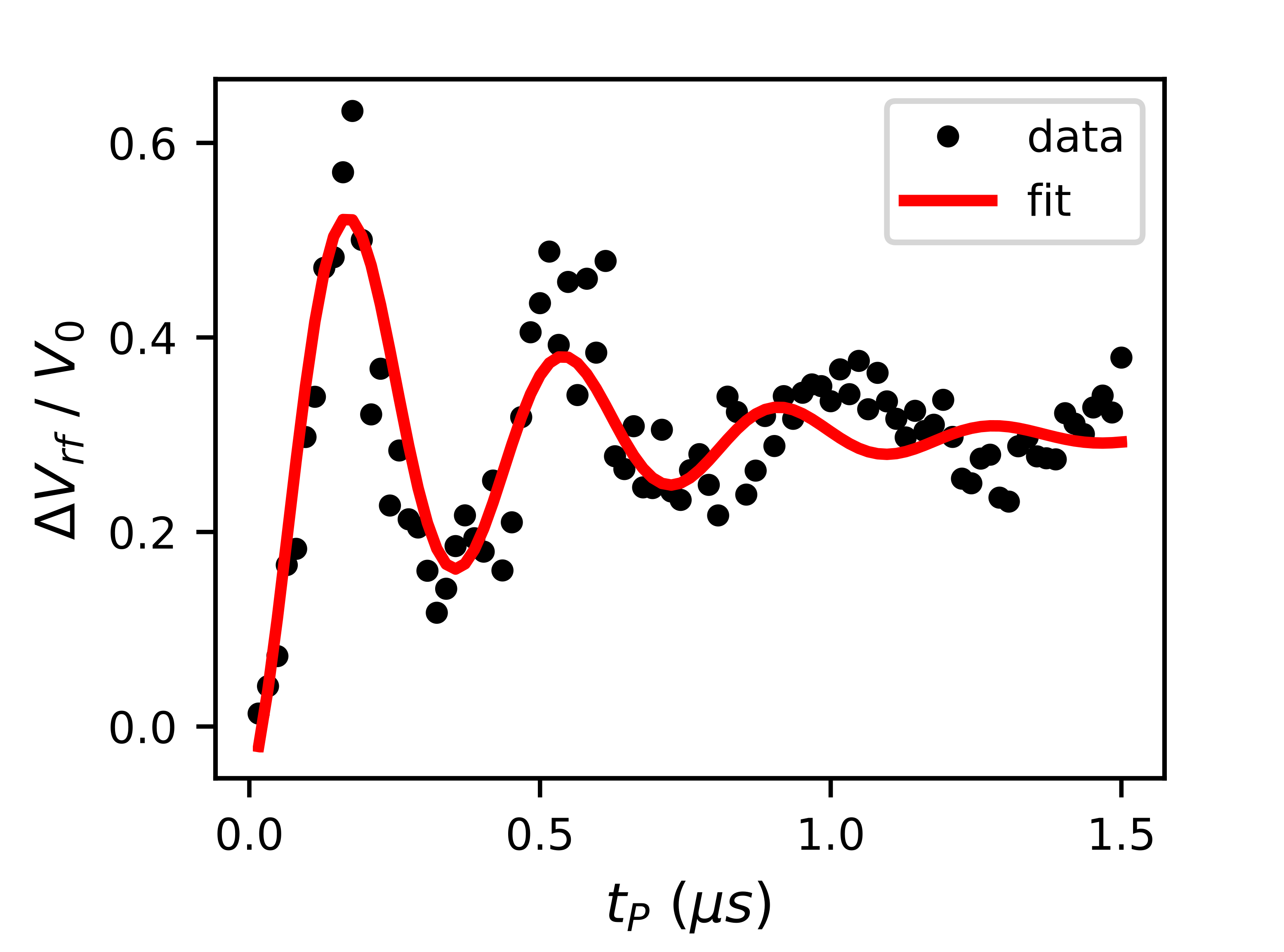}
    \caption{Extraction of dephasing time. Coherent oscillation measured in the singlet-triplet basis fitted to a damp sine wave with $T_2=0.4 \pm 0.1~\upmu$s, $f_{Rabi}=17.0 \pm 0.4$~MHz and $A=0.35 \pm 0.04$ fitted to Eq.~\ref{eq:rabi}.}
    \label{fig:1dcoherent}
\end{figure}

Once at point `P', $\ket{(1,1)S}$ dephases into a mixture of $\ket{(1,1)T_0}$ and $\ket{(1,1)S}$ . We characterise the dephasing time by varying the time $t_P$ at detuning point `P' in the (1,1) region. At this point, a combination of the spin-orbit interaction at the Si:SiO2 interface and the exchange interaction drives $\ket{\uparrow\downarrow}$ and $\ket{\downarrow\uparrow}$ mixing \cite{Jock2018, chittock2024exchange}. We measure Rabi oscillations between the $\ket{ \rm S_0}$ and $\ket{ \rm T_0}$ states as shown in Fig.~\ref{fig:1dcoherent}. We fit the oscillation to a damped sine wave:

\begin{equation}
\label{eq:rabi}
    V_{RF}(t) = A \exp(-t/T_2^*) \sin(f_{Rabi}t + \phi)
\end{equation}
where $A$ is a normalisation factor, $T_2^*=0.4 \pm 0.1~\upmu$s is the dephasing time, $f_{Rabi}$ is the frequency of the oscillation and $\phi$ the phase. Waiting for ~$1\upmu$s, this produces a mixture of states with expected probabilities $p(\ket{\rm S}) \simeq 0.25$, $p(\ket{\rm T_0}) \simeq 0.25$ and $p(\ket{\rm T_-}) \simeq 0.5$.

When performing readout, the tunnelling transition from (1,1) to (0,2) is not observed as it occurs on a timescale much faster than the sensor's minimum integration time, $\tau_{\rm min}$, (defined as the integration time required to distinguish (1,1) from (0,2) with SNR of one), which we measure as $\tau_{\rm min} = 3.3~\upmu\mathrm{s}$ (see Supplementary Note~\ref{app: tmin}).
%
Given the charge tunnelling process is fast, the relaxation rate $\Gamma_{\rm T_-}$ is primarily limited by the spin-flip rate, and $\Gamma_{\rm T_0}$  by spin dephasing. We therefore expect $\Gamma_{\rm T_-}\ll\Gamma_{\rm T_0}$ as spin flips are driven by weaker interactions, such as spin-orbit coupling~\cite{meunier2007experimental, shen2007triplet}. Hence, we assign the shorter of the measured relaxation times to $\Gamma_{\rm T_0}^{-1}=170 \pm 2~\upmu$s, and the longer to $\Gamma_{\rm T_-}^{-1}=290 \pm 3$~ms in the regime of $V_{\rm GTB2}=212~$mV.

\subsection{Fidelity calculation using an analytical model}
\subsubsection{Two-state model}
\label{app: fid_anayltical}

The histogram obtained from repeated single-shot measurements corresponds to two peaks associated with each charge state (1,1) and (0,2) separated by the signal strength $\delta V_{\rm rf}$ and broadened equally by Gaussian noise, which holds when the noise is dominated by the noise temperature of the HEMT amplifier~\cite{derakhshan2020charge}. In the simplest case, we can model the data by assuming the two charge states correspond to two distinct spin states. This causes an asymmetry in the distribution due to relaxation from the excited (triplet) (1,1) state into the singlet (0,2) state, further broadening the triplet peak. Fidelity is calculated by fitting the experimental histogram to this bimodal distribution and finding an optimal threshold voltage that minimises the classification error, i.e. the overlap of the underlying distributions of the two states being discriminated. 

Specifically, the singlet state generates white noise ($\sigma(t)$) centred at the singlet voltage $V_S$. This results in a signal distribution $n_S(V,t)$ at integration time $t = t_\mathrm{read}$, described by the following equation: 

\begin{equation}
\label{eq:Gauss_singlet}
    n_S(V,t) = \frac{1}{\sqrt{2\pi}\sigma(t)} \text{exp}\left(-\frac{(V-V_S)^2}{2\sigma(t)^2}\right).
\end{equation}
In the presence of Gaussian noise, the standard deviation decreases as a function of the integration time $t$. We can write the standard deviation as $\sigma(t) = \sigma_0\sqrt{\frac{t_0}{t}}$, where $\sigma_0$ is the standard deviation at the reference integration time $t_\mathrm{read}=t_0$. 

For a triplet state, the signal remains at the triplet voltage $V_{T}$ for a time $T_1 = \Gamma^{-1}$ before decaying to the singlet state. Here we do not make any assumption on the triplet state at stake, and consider that only one triplet state is to be considered e.g. the $\ket{T_-}$ state for parity readout (considering all $\ket{T_0}$ has relaxed) or the $\ket{T_0}$ state for singlet-triplet readout. The triplet distribution is modelled as:

\begin{equation}
\label{eq:Gauss_triplet}
    n_T(V,t) = \frac{1}{\sqrt{2\pi}\sigma(t)} \text{exp}\left( -\frac{t}{T_1}\right) \text{exp}\left(-\frac{(V-V_T)^2}{2\sigma(t)^2}\right) + I_D,
\end{equation}

where \textbf{$I_D$} represents the tail that forms due to short-lived triplets decaying into a singlet. Such tail is given by~\cite{Barthel2009, Gambetta2007, DAnjou2014}:

\begin{widetext}
\begin{align}
    I_D = &\frac{t}{T_1} \frac{1}{\sqrt{8\pi}(V_T-V_S)} \text{exp} \left[\frac{t}{T_1} \frac{1}{V_T-V_S}\left( V_S-V + \frac{\sigma^2(t)}{2(V_S-V_T)} \frac{t}{T_1} \frac{1}{V_T-V_S}\right) \right] \\
    &\left[ \text{erf}\left( \frac{t}{T_1} \frac{\sigma(t)}{\sqrt{2}(V_S-V_T)} + \frac{V-V_S}{\sqrt{2}\sigma(t)}\right) - \text{erf}\left( \frac{t}{T_1} \frac{\sigma(t)}{\sqrt{2}(V_S-V_T)} + \frac{V-V_T}{\sqrt{2}\sigma(t)}\right)\right] \nonumber
\end{align}
\end{widetext}

\begin{figure}
    \centering
    \includegraphics[width=1.0\linewidth]{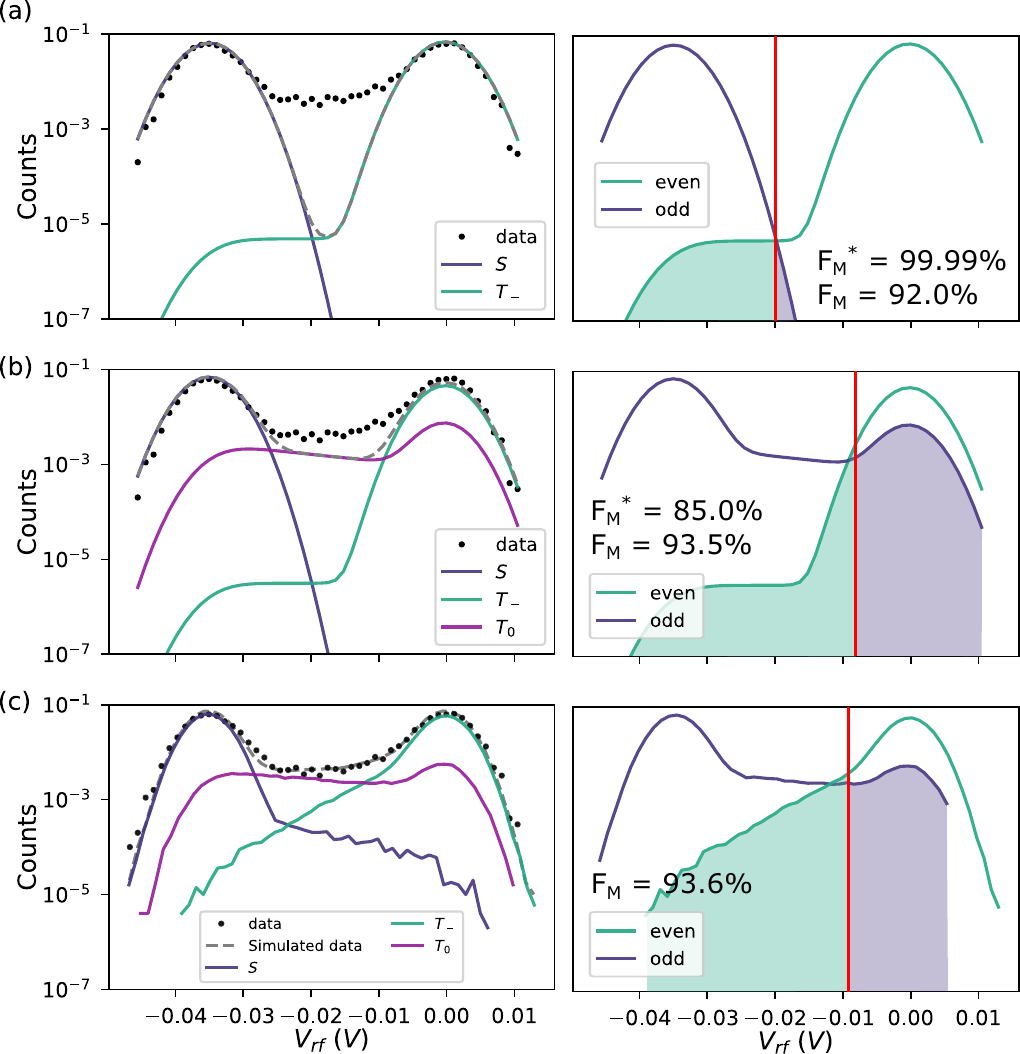}
    \caption{Comparison of data simulation for fidelity calculation. Histogram of 10.000 single-shot data (points) integrated for $t_{\rm read} =204 ~\upmu$s, with $\Gamma_{T_{0}}^{-1}=170~\upmu$s, $\Gamma_{T_{-}}^{-1}=230~$ms, and simulated distribution (line) via three different methods. Histograms are normalised, and the x-axis is the raw RF signal. In the second column, we consider the odd $(\ket{S}, \ket{T_0})$ and even $(\ket{T_{-}})$ distributions and display the optimal threshold voltage which minimises the readout error (shown as the shaded region). $F^*_{\rm m}$ is calculated from the analytical distribution (using Eq.~\ref{eq:fidelity_analytical} for (a), Eq.~\ref{eq:fidelity_parity} for (b)) and $F_{\rm m}$ from the 'real' distribution found in panel (c) (Eq.~\ref{eq:fidelity}). (a) The data is fitted to a two-state model via Eq.~\ref{eq:distribution}. The threshold is optimised for $F_{\mathrm m}^*$. We find that $F_{\mathrm m}^*>F_{\mathrm m}$ and thus the two-state model overestimates greatly the fidelity, neglecting the undecayed odd parity triplets. (b) We extend to a three-state model which accounts for $\ket{T_0}$ states, improving the fit but still not capturing all the data. The optimal threshold voltage is shifted up to minimise the error due to $\ket{T_0}$ classified as even parity. We find (i) $F_{\mathrm m}^*<F_{\mathrm m}$ showing this model underestimates fidelity as it mistakes TLF and accounts for more $\ket{T_0}$ states than there is (ii) $F_{\mathrm m}$ is larger than in panel (a) because it corresponds to a more optimal threshold. (c) The data is simulated using a HMM that contains $\ket{T_0}$ and the TLF, and captures the data well. We note that classifying the single-shot traces with the HMM and not with the threshold method gives an improved fidelity of $F_{\rm m}=95.9\%$ for that same integration time.  }
    \label{fig:histogram_barthel}
\end{figure}

Equations ~\ref{eq:Gauss_singlet} and~\ref{eq:Gauss_triplet} allow us to identify the parts of the histogram corresponding to each spin state. We can then fit the raw data histogram to the combined distribution:

\begin{equation}
\label{eq:distribution}
    n(V,t) = P_T\cdot n_T(V,t) + P_S\cdot n_S(V,t),
\end{equation}

where $P_S$ and $P_T$ are the fractions of traces in the singlet and triplet states, respectively, with $P_S + P_T = 1$. Figure~\ref{fig:histogram_barthel}(a) shows the fit of Eq.\ref{eq:distribution} to the histogram of single-shot data.

Using the analytical formula, fidelities for the singlet state $F^*_S(V_{th})$ and the triplet state $F^*_T(V_{th})$ can be calculated as the fraction of events misclassified given the threshold voltage $V_{th}$:

\begin{align}
F^*_S(V_{th}) &= 1 - \int_{V_{th}}^{\infty} n_S(V) \, dV \; \text{and} \\
F^*_T(V_{th}) &= 1 - \int_{-\infty}^{V_{th}} n_T(V) \, dV  \\ \nonumber
\end{align}

The fidelity metric assumes equal probabilities for singlet and triplet states ($P_T = P_S = 0.5$), treating both types of misclassification equally. Among these metrics, visibility $V^*_m$ is defined as:

\begin{equation}
\label{eq:visibility_analytical}
  V^*_m = \max\limits_{V_{th}}\left( F_S^*(V_{th}) + F_T^*(V_{th}) - 1\right),  
\end{equation}

which can take values between zero and one.

Alternatively, the measured fidelity $F^*_{\mathrm m}$ determines the average error in assigning the wrong label to the measured state and thus takes values between 0.5 and 1:

\begin{equation}
\label{eq:fidelity_analytical}
    F^*_{\mathrm m} = \max\limits_{V_{th}} \left(\frac{F^*_S(V_{th}) + F^*_T(V_{th})}{2}\right).
\end{equation}

This is the metric usually quoted in the literature as $F_{\mathrm m}$.  The maximum measured fidelity will occur when $n_S(V_{th}) = n_T(V_{th})$, which means that we can also define $F_{\mathrm m}^*$ as:
 \begin{equation}
 \label{eq:fidelity_Vth}
     F_{\mathrm m}^* = \frac{1}{2} \left[ 1 - \text{erf} \left( \frac{V_S-V_{th}}{\sqrt{2} \, \sigma} \right)\right]
 \end{equation}
In Fig.~\ref{fig:histogram_barthel}, we visualise the parity readout error by the shaded area which corresponds to the integral of the even (respectively odd) density below (above) threshold, coloured in green (purple). In panel (a), we calculate a spin parity measurement fidelity of 99.99\%, however, this value is likely overestimated, as indicated by the mismatch between the data and the fitted histograms. The intersection of the peaks in the model is smaller than observed in the experimental data, which can be attributed to the presence of $\ket{T_0}$ spin states and a two-level fluctuator, factors not accounted for and leading to an inflated fidelity estimate.

\subsubsection{Electrical fidelity}
 
We also note the definition of the electrical fidelity, which corresponds to the trivial case for which the relaxation time is taken as infinite and thus $V_{th}$ will be the average between $V_S$ and $V_T$.
This gives an expression for the electrical fidelity $F_E$ that does not account for spin-relaxation errors:

\begin{align}
    F_E^* &= \lim \limits_{T_1\rightarrow \infty} F_{\mathrm m}^* = \frac{1}{2} \left[ 1 + \text{erf} \left( \frac{V_T-V_S}{2 \sqrt{2} \, \sigma} \right)\right] \\ \nonumber
    &=  \frac{1}{2} \left[ 1 + \text{erf} \left( \frac{\text{SNR}}{2 \sqrt{2}} \right)\right].
\end{align}

The electrical fidelity is useful for discerning the nature of the readout errors. $1-F_E^*$ gives the errors due to the sensor, while $F_E^*-F^*_{\mathrm m}$ are the relaxation errors.

\subsubsection{Three-state model}
To refine the fit of Eq.~\ref{eq:distribution} and account for undecayed $\ket{T_0}$ spin states, we can add a third distribution $n_{T_0}$ with probability $P_{T_0}$ where  $P_S + P_{T_-} + P_{T_0}=1$. For parity readout, we now define 

\begin{align}
F^*_{\rm odd}(V_{th}) &= 1-\int_{-\infty}^{V_{th}}(n_S+n_{T_0}) \ dV \; \text{and} \\
F^*_{\rm even}(V_{th}) &= 1-\int_{V_{th}}^{+\infty}n_{T_-}\ dV
\end{align}
and therefore, measurement fidelity is calculated as 
 \begin{equation}
 \label{eq:fidelity_parity}
 F^*_{\mathrm m} = \max\limits_{V_{th}} \left( \frac{F^*_{\rm odd}(V_{th}) +F^*_{\rm even}(V_{th})}{2}\right)
 \end{equation}
To avoid biasing classification, we fix $P_S=0.25, \ P_{T_0}=0.25, \ P_{T_-}=0.5$ in the case of parity readout.

We calculate the density $n_{T_0}$ similarly to that of Eq.~\ref{eq:Gauss_triplet}, changing to the appropriate relaxation time. When fitting the data to this overall distribution, we notice a subsequent improvement to the fit quality as seen in Fig.~\ref{fig:histogram_barthel}(b). This is expected since the histogram is measured at $t_{\rm read} =204 ~\upmu$s, a value comparable to $\Gamma_{T_0}^{-1} = 168 ~\upmu$s, therefore leaving a significant number of undecayed $T_0$ states. However, the three-state model is still insufficient to fully capture the data, especially between the two peaks. As we explain in the next section, this is due to the effect of a two-level-fluctuator visible in the shoulder of the singlet distribution towards increasing $V_{\rm rf}$.

\subsection{Fidelity calculation with simulated data using a HMM}
We show the presence of the two-level-fluctuator in Supplementary Information~\ref{app: data_simulation}. We now simulate each density $n_i$ with a Hidden Markov Model (HMM) that incorporates six hidden states: the three spin states and the corresponding ones with two-level-fluctuators. The transitions between the hidden states follow a memoryless process, where the triplet states ($\ket{ \rm T_0}$ and $\ket{ \rm T_-}$) decay to the singlet state $\ket{ \rm S}$ at their measured relaxation rates. Each hidden state emits a constant signal with Gaussian noise, corresponding to the sensor value for the different charge states: (1,1) for the triplets and (0,2) for the singlet. The standard deviation of the emissions reflects the experimental noise in each state. 

The HMM is fitted to the data by first simulating many single-shot traces, averaging them over $t_{\rm read}$, and generating a model probability distribution. In that process, the parameters for the HMM are updated using the Expectation-Maximization algorithm~\cite{dynamaxlibrary}, which iteratively refines the parameters (initial state, transition, and emission probabilities) to maximise the likelihood of observing the data given the model parameters. The overall distribution fully captures the system's noise, as we show in Fig.~\ref{fig:histogram_barthel}(c).

Having obtained new underlying probability distributions, fidelity is calculated for each spin state as:
\begin{align}
F_S &= 1 - \frac{\text{Singlets misclassified}}{\text{Total number of predictions}}, \; \text{and} \\ \\ \nonumber
F_T &= 1 - \frac{\text{Triplets misclassified}}{\text{Total number of predictions}}. \\ \nonumber
\end{align}

The overall fidelity is then calculated as the average fidelity across all spin states:

\begin{equation}
\label{eq:fidelity}
F_{\mathrm m} = \frac{1}{N} \sum_{i=1}^N F_i.
\end{equation}

For two spin states, this simplifies to:

\begin{equation}
\label{eq:fidelity}
F_{\mathrm m} =1 - \frac{\text{Number of errors}}{2 \cdot (\text{Total number of predictions})}.
\end{equation}

Similarly to before, such fidelity is calculated at each threshold voltage and the overlap of the states being discriminated (e.g.~odd/even parity) is minimised by setting the optimum threshold value. 

Here, we note although fidelity is widely used, it has significant limitations. For example, when all classifications are systematically incorrect, the minimum value of $F_{\mathrm m}$ depends on the number of states. With two spin states, the minimum fidelity is $F_{\mathrm m} = 0.5,$ while for three states, it is $F_{\mathrm m} = 0.66$. This inconsistency makes fidelity a poor metric for systems with more than two states.

Visibility addresses these shortcomings. Defined analogously to accuracy in machine learning, visibility measures the proportion of correctly classified events:

\begin{equation}
\label{eq:Visibility}
V_\mathrm{m} = 1- \frac{\text{Number of errors}}{\text{Total Number of Predictions}}.
\end{equation}

Unlike fidelity, visibility is consistent across systems with any number of spin states and can naturally extend to applications involving qudits or other multi-state systems~\cite{chen2023transmon, wang2024systematic}. This makes it a more reliable and versatile metric for evaluating classification performance. However, we have used fidelity as our metric for consistency and comparison with the literature.

\subsection{Classification methods}

In this work, we use two methods to classify readout traces: the widely used threshold technique and a more advanced approach based on a HMM.

\label{app: classification}
\subsubsection{Threshold}
The threshold method is a simple and widely used technique for labelling spin states. The signal is averaged over a set integration time ($t_\mathrm{read}$), and the spin state is determined by comparing this average to a predefined threshold~\cite{Barthel2009}. This is the threshold we refer to when calculating readout fidelity.

\subsubsection{Hidden Markov Model (HMM)}

In the main text and when calculating fidelity, we describe how a Gaussian Hidden Markov Model (HMM) can represent the dynamics of our system. A HMM can model the hidden states (e.g. spin configurations) and relate them to the observed sensor signal.

The forward-backwards algorithm calculates the likelihood that the system is in a specific hidden state at any time $t$, based on the observed signal up to $t$ and also data measured after $t$, up to the readout time. This probability is given by:

\begin{equation}
    p(z_\mathrm{t}| y_\mathrm{1:T}, \theta),
\end{equation}

where:

\begin{itemize}

    \item $z_\mathrm{t}$: hidden state at time $t$,
    
    \item $y_\mathrm{1:T}$: observed signal over the entire measurement,

    \item $\theta$: the HMM parameters, including initial state probabilities, the transition matrix, and emission probabilities.
\end{itemize}

For classifying spin states, we use the forward-backwards algorithm to determine the most likely hidden state at the start of the measurement (t = 0). This is achieved by maximizing the probability:

\begin{equation}
\text{predicted state} = \arg\max_{z_0} \, p(z_0 \mid y_{1:T}, \theta)
\end{equation}

Unlike the threshold method, this approach takes advantage of the entire time-dependent structure of the observed data collected during the integration period ($t=t_\mathrm{read}$), providing improved classification accuracy.

We implement this analysis using the Python library \texttt{dynamax}~\cite{dynamaxlibrary}, which offers efficient tools for HMM parameter estimation and state inference\cite{murphy2012probabilistic}. The data simulation for our system that incorporates the three spin states and a two-level fluctuator is detailed in Supplementary Information~\ref{app: data_simulation}.

\begin{acknowledgments}
\noindent We acknowledge helpful conversations with S.C.~Benjamin and G. Burkard at Quantum Motion. We also acknowledge technical support from J. Warren. This work received support from the European Union's Horizon 2020 research and innovation programme under grant agreement No. 951852 (Quantum Large Scale Integration in Silicon); from the Engineering and Physical Sciences Research Council (EPSRC) under grant Nos.~(EP/S021582/1), (EP/L015978/1), (EP/T001062/1), and (EP/L015242/1); and from Innovate UK under grant Nos. (43942) and (10015036). M.F.G.Z.~acknowledges support from the UKRI Future Leaders Fellowship (MR/V023284/1). 

\end{acknowledgments}

\appendix

\section{Spin readout fidelity benchmark}
\label{app:fid_benchmark}
We gather reported spin readout fidelities in semiconductor nanostructures from the literature and summarise them in Table \ref{tab:fidelity}. Our focus is on singlet-triplet and parity readout measured in single-shot via Pauli spin blockade (PSB) across platforms, including semiconductor heterostructures (e.g. Si/SiGe) and silicon devices (both MOS and nanowire (NW) structures). Important metrics are fidelity $F_{\mathrm m}^*$, integration time $t_{\rm read}$ and relaxation rate $\Gamma$ of the slowest triplet depending on the readout basis. In all these works, fidelity $F_{\mathrm m}^*$ is calculated using the threshold method and bimodal fitting of histograms. 

\begin{table}[]
    \centering
    \begin{tabular}{c|c|c|c|c|c}
         Sensor & Platform & $F^*_{\mathrm m}$ & $t_{\rm read}$ & $\Gamma^{-1}$ & Ref. \\
         \hline
          SET & MOS & 99.9 & 100 $\mu$s & undisclosed & \cite{steinacker2024300} \\ 
        RF-SET & Si/SiGe & 99.9 & 2 $\mu$s  & 18.2 ms & \cite{takeda2024rapid} \\ 
        \hline
        SEB & NW & 99.2 & 6 $\mu$s & 230 $\mu$s & \cite{oakes2023fast} \\
        SEB  & NW & 99.9 & 20 $\mu$s & 32 ms& \cite{Niegemann2022} \\

    \end{tabular}
    \caption{Overview of readout performance across various using different sensors.}
    \label{tab:fidelity}
\end{table}

\section{Lever arm extraction}
\label{app: leverarm}
\begin{figure}
    \centering
    \includegraphics[width=0.8\linewidth]{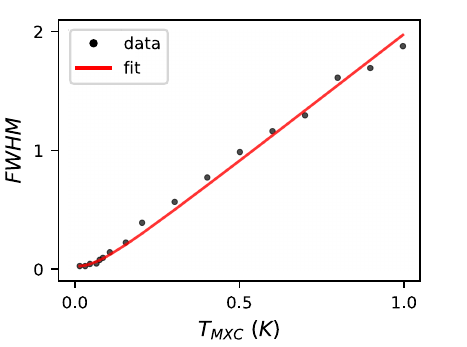}
    \caption{SEB thermometry. Measurement of the SEB FWHM as a function of temperature and fit Eq. \ref{eq:thermo} to extract $T_e = 90 \pm 11$mK and $\alpha=0.17 \pm 0.01$.}
    \label{fig: thermometry}
\end{figure}

We use Coulomb peak thermometry to find the SEB lever arm and electron temperature. We use a SEB DRT peak that is nor power or lifetime broadened. We fit the FWHM of the peak to \cite{Ahmed2018} 
\begin{equation}
\label{eq:thermo}
    \rm FWHM = \frac{3.53k_B}{e \alpha} \sqrt{T_{\text{MXC}}^2 + T_e^2}
\end{equation}
where $\alpha$ is the gate lever arm, $\rm T_{\text{MXC}}$ is the mixing chamber temperature and $\rm T_e$ is the electron temperature. Fig.~\ref{fig: thermometry}  shows the fit where we extract $\rm T_e = 90 \pm 11$~mK and $\alpha=0.17 \pm 0.01$.

\section{Data Pre-Processing for Readout}
\label{app: pre-processing}

\begin{figure}
    \centering
    \includegraphics[width=\linewidth]{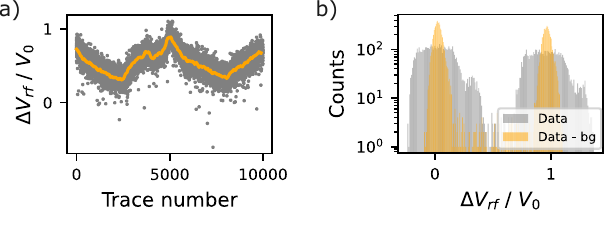}
    \caption{Data Pre-Processing for Readout. a) Sensor drift observed during the readout fidelity process (shown in grey), whose fluctuations are mitigated by applying a moving average filter with a window size of 50, as shown in orange. The interval between consecutive readout traces depends on the duration of the entire readout sequence, which includes an initialization phase requiring a wait time of approximately five times $T_1$. This example corresponds to the longest  $T_1$, resulting in the longest readout sequence, lasting nearly one second. (b) Histograms comparing 10,000 single-shot traces averaged over $t_\mathrm{int}=204$~$\mathrm{\mu s}$ before (grey) and after (orange) background effect removal.}
\label{fig:app_sensor_drift}
\end{figure}

In this section, we address the issue of sensor drift, which degrades readout fidelity. To mitigate this problem, we acquire data during the final 120~$\mathrm{\upmu s}$ of the plunge (`P') phase in the voltage sequence when the system is in the known (1,1) charge configuration. This allows capturing the background noise before actual data collection begins. However, the background signal is susceptible to noise, as shown by the grey data points in Fig.~\ref{fig:app_sensor_drift}a. To reduce this noise and stabilise the readout, we average the background from the last 50 readout traces. This averaged value is subtracted from each readout trace to correct for the sensor drift. 

Importantly, we only use background measurements of traces taken previously (in the past), ensuring that this approach remains compatible with real-time quantum algorithm operations. This is crucial because, during the execution of a quantum algorithm, decisions and corrections must rely on data that has already been collected; future measurements are not available in real time.

The effectiveness of this method is demonstrated in Fig.\ref{fig:app_sensor_drift}a, where the orange line represents the averaged background over time. After applying the correction, the sensor signal no longer drifts, resulting in a more stable output. Further evidence is shown in Fig.\ref{fig:app_sensor_drift}b, where the histogram of averaged single-shot traces over the first 204~$\upmu$s reveals a reduced overlap between peaks after drift removal, indicating enhanced signal discrimination.

This method is compatible with quantum algorithms, as periodic background measurements could be performed during quantum operations, provided that the sensor-rf signal does not significantly affect qubit gate performance. Continuously tracking and compensating for sensor drift ensures reliable and high-fidelity readout throughout dynamic quantum computations.

\section{Data Simulation with a HMM}
\label{app: data_simulation}
In the main text, we present a Gaussian Hidden Markov Model (HMM) that model three hidden states: $\ket{ \rm S}$, $\ket{ \rm T_-}$, and $\ket{ \rm T_0}$ and captures the influence of a two-level fluctuator (TLF) that alternates between a ground (G) and an excited (E) state, introducing additional noise (see Fig.\ref{fig:histogram_barthel}).

\subsection{TLF Effect on Sensor Signal}

\begin{figure}
    \centering
    \includegraphics[width=\linewidth]{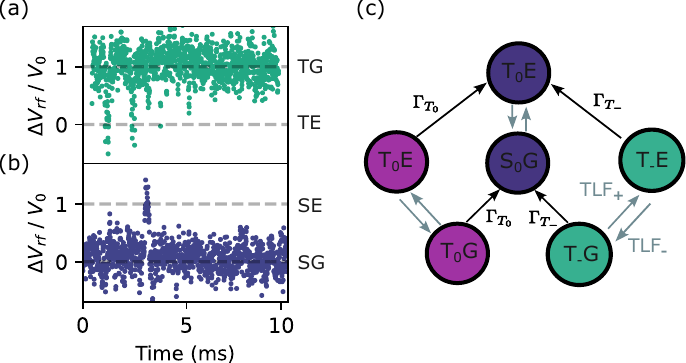}
    \caption{TLF effect on Sensor Signal. a) Effect of the two-level fluctuator (TLF) on the sensor signal for the singlet state ($ \ket{ \rm S} $).  When the TLF is in the ground state (G), the sensor signal is centred at  $\Delta V_\mathrm{rf} / V_0 = 0$. However, when the TLF switches to the excited state (E), the signal shifts to resemble a triplet state ( $\Delta V_\mathrm{rf} / V_0 = 1$ ). Effect of the two-level fluctuator (TLF) on the sensor signal for the triplet states ($ \ket{ \rm T_-} $). c) Markov chain representing the six hidden states, including the transitions of the two-level fluctuator:  $\text{TLF}_ +$  for transitions from the ground (G) to the excited (E) state, and  $\text{TLF}_-$  for transitions from the excited (E) to the ground (G) state.}
\label{fig:app_TLF}
\end{figure}

The TLF alters the sensor’s output depending on its state. For $\ket{ \rm S}$, the sensor emits a signal centred at $\rm \Delta V_{rf} / V_0 = 0$ when the TLF is in the ground state (G). However, when the TLF switches to the excited state (E), the signal shifts dramatically to resemble a triplet signal ($\rm \Delta V_{rf} / V_0 = 1$), as illustrated in Fig.~\ref{fig:app_TLF}a.

A similar phenomenon occurs for $\ket{ \rm T_-}$ and $\ket{ \rm T_0}$, where the signals, typically centred at $\rm \Delta V_{rf} / V_0 = 1$ in the ground TLF state, shift to resemble a singlet signal ($\rm \Delta V_{rf} / V_0 = 0$) when the TLF is excited (Fig.~\ref{fig:app_TLF}b).

\subsection{Extended HMM with TLF States}

To account for this effect, the HMM is expanded to include six states: $\ket{ \rm S}$G, $\ket{ \rm S}$E, $\ket{ \rm T_-}$G, $\ket{ \rm T_-}$E, $\ket{ \rm T_0}$G, and $\ket{ \rm T_0}$E. These states incorporate both spin and TLF dynamics (see Fig.~\ref{fig:app_TLF}c). Transitions follow two mechanisms:

\begin{itemize}
    \item 	\textbf{Spin State Transitions:} Triplet states ($\ket{ \rm T_-}$, $\ket{ \rm T_0}$)  decay to $\ket{ \rm S}$, preserving the TLF state. For example, $\ket{ \rm T_0}$G decays to $\ket{ \rm S}$G, and $\ket{ \rm T_0}$E decays to $\ket{ \rm S}$E. 
    
    \item 	\textbf{TLF Switching:} Any state can transition between its G and E variants.

\end{itemize}

This extended HMM better captures the system dynamics and noise effects, aligning closely with observed results.

\section{Single-shot SNR optimisation}
\label{app: RF opti}
For reflectometry readout, the voltage SNR can be given as \cite{Vigneau2023}
\begin{equation}
\label{eq:power}
    \rm SNR = |\Delta \Gamma_{c.s.}| \frac{V_{in}}{V_n},
\end{equation}
where $V_{in(n)}$ are the input RF and noise voltages, respectively and $\Delta \Gamma_{c.s.}$ is the change of signal due to a charge sensing event. For the SEB a DRT transition will cause a change of quantum capacitance given by Eq.~\ref{eq:cap_seb}. In the regime where this change is small\footnote{We are referring to the small-signal regime where $Q_r \Delta C_{DRT} / C_{tot} << 1$} it causes a change of signal \cite{ciriano2022thesis, oakes2024novel} :
\begin{equation}
    \Delta \Gamma = \Delta C_{\rm DRT} \left[ \frac{\partial \Gamma}{\partial C_{\rm DRT}} \right]_{f=f_{\rm res}}
\end{equation}
The second term of the equation can be calculated as a function of the resonant frequency and therefore the tank circuit parameters such that :
\begin{equation}
\label{eq:snrtankcircuit}
    \Delta \Gamma = i \frac{2\beta}{(1+\beta)^2} Q_{int} \frac{\Delta C_{\rm DRT}}{C_{\rm tot}} 
\end{equation}
where $C_{\rm tot}$ is the total capacitance of the system (including the SEB and matching network), and $\Delta C_{\rm DRT}$ is the change in capacitance of the SEB as defined in the main text.   
Finally, we have to account that the signal measured for charge sensing does not correspond to the full height of the SEB peak. This is because we aim at differentiating peaks $(0,2)$ and $(1,1)$ and what we therefore measure is the signal contrast $\Delta V_{RF}$ as detailed in the main text and shown in Fig.~\ref{fig:fig1}(c). To account for this reduction in signal we introduce the parameter $\eta$ such that:
\begin{equation}
    \label{eq:snr_eta}
    \Delta \Gamma_{\text{c.s.}} =  \eta \Delta \Gamma,
\end{equation}
where $\Delta \Gamma_{\text{c.s.}}$ is the change in signal due to a charge sensing event. Combining Eqs.\ref{eq:power}, \ref{eq:snrtankcircuit}, \ref{eq:snr_eta} yield 
\begin{equation}
\label{eq:snrtotal}
    \rm SNR = \eta \frac{2\beta}{(1+\beta)^2} Q_{int} \frac{\Delta C_{\rm DRT}}{C_{\rm tot}} \frac{V_{in}}{V_n} .
\end{equation}

Therefore, we see that maximising the SNR entails the optimisation of  1) the input amplitude or power via Eq.~\ref{eq:power} 2) the tank circuit parameters via Eq.~\ref{eq:snrtankcircuit} and 3) the DRT transition via  Eq.~\ref{eq:cap_seb}. Power is optimised by increasing it until the SEB charge sensor peak gets broadened leading to a reduction of $\Delta C_D$ 
that dominates the SNR. We mitigate the back-action of the RF on relaxation times by tuning the SEB such that it is on resonance (peak) when the qubits are in the (0,2) ground state. 
The optimal power used is $P_{in}$ = -80~dB. The tank circuit components are cleverly chosen as explained in Supplementary Note~\ref{app: set-up}. Optimisation of the DRT is the focus of the main text.

\section{Noise spectrum}
\label{app:snr_chargenoise}
\begin{figure}
    \centering
    \includegraphics[width=0.8\linewidth]{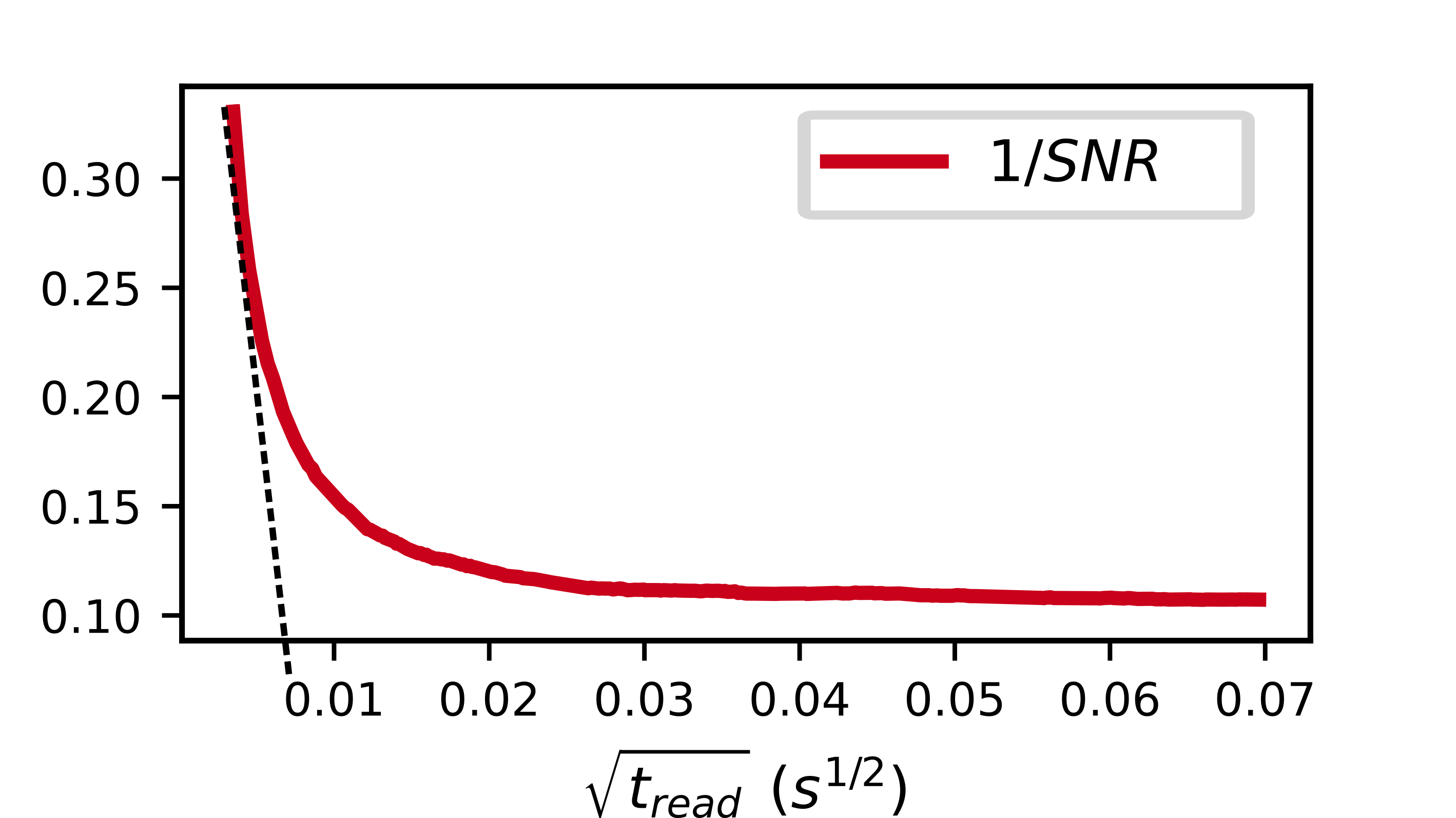}
    \caption{Noise analysis. At fixed signal 1/SNR = $\sigma$ which is suppressed linearly with $1/\sqrt{t_{\rm read}}$ up until it plateau for $t_{\rm read}^{-1} \sim 30$kHz. This is where non-white noise such as charge noise starts to dominate.}
    \label{fig:1/snr}
\end{figure}

We confirm the noise is dominated by white noise at short integration time by plotting 1/SNR with SNR obtained from Eq.\ref{eq:fidelity} in the case of $\Gamma=0$ (electrical fidelity). We use $F_e^*$ measured for the best tuning point of Fig.~\ref{fig:fig3} i.e. $V_{\rm GBB1} = 227~$mV. This is shown in Fig.~\ref{fig:1/snr}. 

\section{SEB dot for sensing}
\label{app: elecnum}

The SEB array is tuned in the many-electrons regime using the plunger gate BP1 and the barrier gate BB1. A stability map showing several DRT transitions is shown in Fig. 7(a). For a barrier gate voltage in the range 0.2-0.3~V, the first electron is loaded into the SEB dot from voltage $V_{GBP1} \simeq 0.5$~V. Given the visible loading voltage of $\Delta V \simeq 33$~mV, we can deduce an approximate electron number of 15-21 from the stability map in Fig. 7(a).

The optimal SNR regime shown in Fig. 3(a) of the main text corresponds to a small range of barrier gate voltages, as shown by the dashed square of the stability map. Beyond the lifetime broadened regime, a large BB1 voltage causes the formation of an additional dot under BB1, which couples strongly to the SEB dot, as hinted by the change in slope above 0.3~V in barrier gate voltage. 
\begin{figure}
    \centering
    \includegraphics[width=0.4\textwidth]{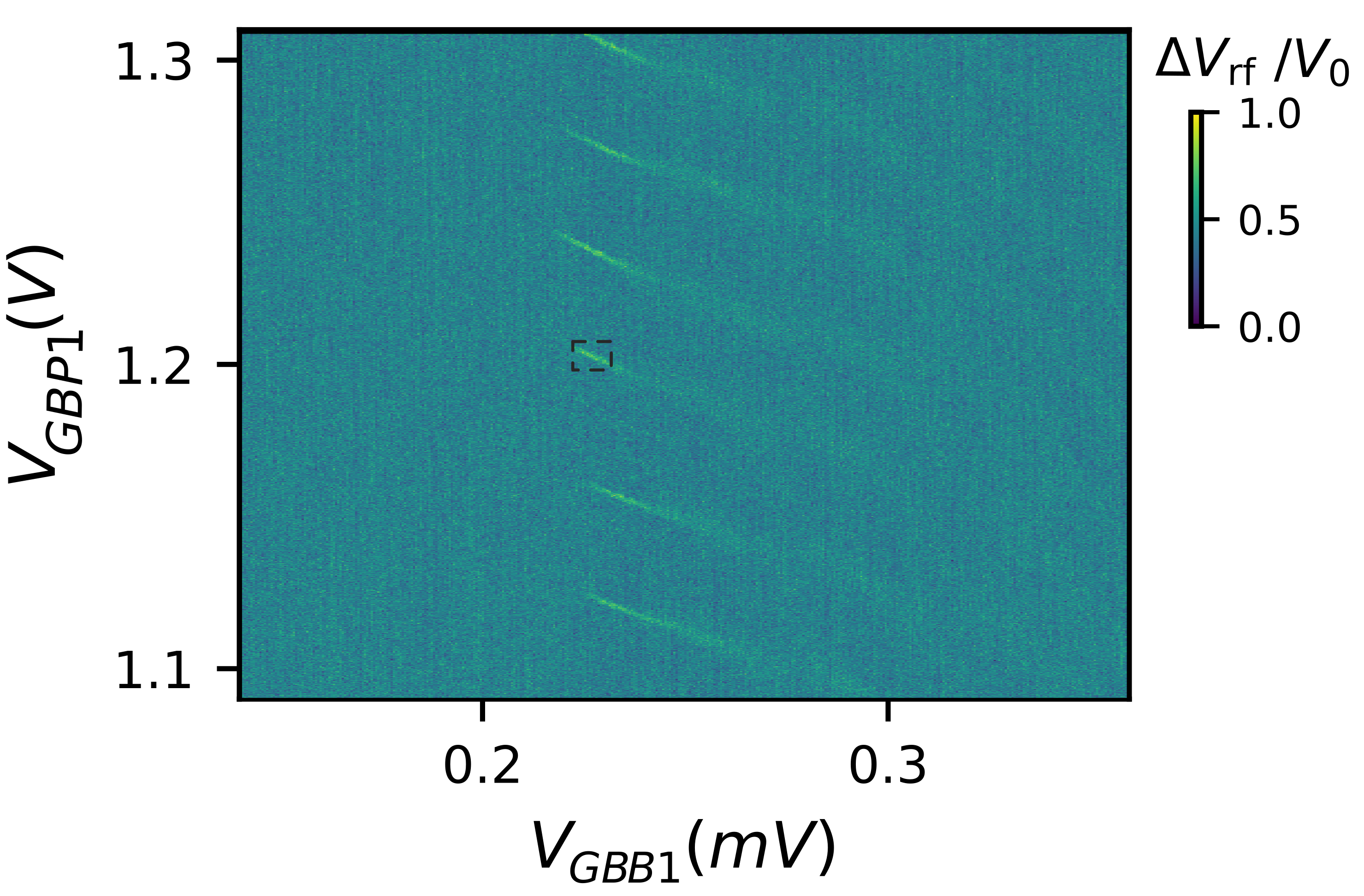}
    \caption{Stability maps of the SEB dot. Six DRT transitions of the SEB  with the lowest electron number around 15 electrons. The dashed rectangle corresponds to the range used for SNR optimisation in Fig.~\ref{fig:fig3}. The data corresponds to two different thermal cycles, causing some voltage shifts.}
    \label{fig:stabmap}
\end{figure}

\section{Tunnel coupling extraction}
\label{app: tunnelcoupling}

\begin{figure}
    \centering
    \includegraphics[width = \linewidth]{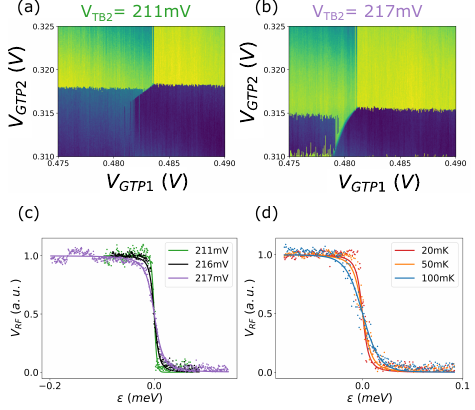}
\caption{Tunnel coupling extraction. (a-b) ICT between (0,2) and (1,1) at two different barrier gate voltage. (c-d) Charge transition across the ICT with fits to Eq. \ref{eq:ictfits} as a function of barrier gate voltages (c) and MXC temperature (d).  }
    \label{fig: t_c}
\end{figure}

In Fig. \ref{fig:fig4} of the main text, we measure tunnel coupling value $t_c$ by using the ICT linewidth and fitting it to \cite{Petta2004}:
\begin{equation}
    \label{eq:ictfits}
    f(\epsilon)=\frac{1}{2}\left[1-\frac{\epsilon}{\sqrt{\epsilon^2+4 t_c^2}} \tanh \left(\frac{\sqrt{\epsilon^2+4 t_c^2}}{2 k_B T_e}\right)\right]
\end{equation}
where $\epsilon$ is the detuning in eV and $T_e$ the qubit electron temperature. The dependency of the ICT on tunnel coupling is visible in Fig.~\ref{fig: t_c}(a-b) where the ICT is shown for two different barrier gate voltages. Fits to equ.\ref{eq:ictfits} are shown in Fig.~\ref{fig: t_c}(c) which determine an upper bound for tunnel coupling value of $t_c = 8.0 \pm 0.5$~$\upmu$eV in the regime where $V_{\rm GTB2} = 217$~mV. Values smaller than $t_c = 4.0 \pm 0.1$~$\upmu$eV however are not accessible as the transition is temperature broadened at around $T_e = 40 \pm 10$~mK, also extracted from the fits obtained when varying the mixing chamber (MXC) temperature at a fixed barrier gate voltage as shown in Fig.~\ref{fig: t_c}(d). We notice the qubit electron temperature is lower than that of the SEB, which could be explained by power dissipation of the RF connected to the SEB.

\section{Minimum integration time}
\label{app: tmin}
\begin{figure}
    \centering
    \includegraphics[width=0.8\linewidth]{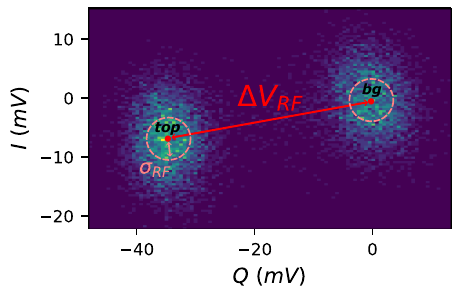}
    \caption{I-Q histogram for 10,000 data traces measured by initialising a 50-50 distribution of $\ket{ \rm S}, \ket{ \rm T_-}$ states.}
    \label{fig:SNR}
\end{figure}

Before projecting the RF signal on the principal component axis, we first examine the single-shot histogram of parity readout in the (I,Q) plane. This is illustrated in Fig.~\ref{fig:SNR}, which corresponds to the measurement shown in Fig.\ref{fig:fig2}(c). In this figure, the left circle represents the even parity charge state signal, whereas the right circle corresponds to the odd parity signal. The sensor is tuned such that the Coulomb peak is on resonance when the qubit is in the even parity state, meaning the left circle approaches the top of the Coulomb peak, whereas the right circle represents the background signal. 

To optimise the signal-to-noise ratio (SNR), we project the signal onto the axis of maximum visibility that connects the centre of the two circles. This axis is referred to as the $\Delta V_{RF}$ axis and is used throughout the text.

We can calculate the SNR of such a measurement as :
\begin{equation}
    \text{SNR} = \frac{\Delta V_{RF}}{\sigma_{RF}},
\end{equation}
where $\Delta V_{RF}$ is the distance between the centres of the two circles and $\sigma_{RF}$ is the standard deviation of the signal distribution. Both parameters are obtained by fitting the histogram to a 2D bimodal Gaussian distribution. The SNR calculated for the data in Fig. \ref{fig:SNR}, is SNR = 8.0 $\pm$ 0.1, averaged over a readout time of $t_{\rm read} = 328 \upmu$s.

Since the histogram reflects the spin signal, the SNR calculated here is for spin readout. However, we can also determine the SNR for the SEB charge sensor, which is slightly higher than the spin readout SNR. This difference arises because spin readout must differentiate between the (0,2) and (1,1) charge states, and the contrast between these states is smaller than the peak height, as seen in Fig. 1(c).

To account for this, we introduce the parameter $\eta$, which represents the fractional change in the signal due to a charge sensing event:$\text{SNR}_{\text{charge}} = \text{SNR}_{\text{spin}}/ \eta$. With $\eta= 0.8$ calculated from Fig. \ref{fig:fig1}(c) we measure an SNR of 10.0 $\pm$ 0.1 for charge readout using the SEB, as shown in Fig. \ref{fig:SNR}.

We then calculate the minimum integration time of the sensor $\tau_{\rm min}$, which is the integration time yielding a power SNR of 1. $\tau_{\rm min}$ is given by \cite{ciriano2022thesis}:
\begin{equation}
    \tau_{\rm min}= t_\text{read}/ \text{SNR}^2.
\end{equation}

We find a minimum integration time for the SEB of $\tau_{\rm min} = 3.3 \pm 0.1 \upmu$s.

\bibliographystyle{apsrev4-2}
\bibliography{apssamp}

\end{document}